\documentclass[a4paper]{article}

\usepackage{anysize}
\marginsize{2.3cm}{2.3cm}{2.3cm}{2.3cm}

\usepackage[table]{xcolor}
\usepackage{graphicx,amssymb,verbatim,epsf,setspace,float,dsfont, soul}
\usepackage[numbers, comma, round, sort&compress]{natbib}
\usepackage{amsmath,amsthm,afterpage}
\newcolumntype{C}[1]{>{\flushright\let\newline\\\arraybackslash\hspace{0pt}}m{#1}}

\newtheorem{proposition}{Proposition}

\makeatletter 
\renewcommand{\thefigure}{S\@arabic\c@figure}
\makeatother

\usepackage{tabularx}

\newcommand{\C}{\mathbb{C}}
\newcommand{\R}{\mathbb{R}}

\newcommand{\cM}{\mathcal{M}}
\newcommand{\cD}{\mathcal{D}}

\newcommand{\V}{\mathcal{V}}
\begin{document}
\title{Numerical algebraic geometry for model selection and its application to the life sciences}

\author{Elizabeth Gross$^{1}$ Brent Davis$^{2}$, Kenneth L. Ho$^{3}$, Daniel J. Bates$^{2}$, Heather A. Harrington$^{4}$\\
$^{1}${Department of Mathematics, San Jos\'{e} State University},\\
$^{2}${Department of Mathematics, Colorado State University},\\
$^{3}${Department of Mathematics, Stanford University},\\
$^{4}${Mathematical Institute, University of Oxford}}
%

\maketitle 
\begin{abstract}
Researchers working with mathematical models are often confronted by the related problems of parameter estimation, model validation, and model selection. These are all optimization problems, well-known to be challenging due to non-linearity, non-convexity and multiple local optima.  Furthermore, the challenges are compounded when only partial data is available.  Here, we consider polynomial models (e.g., mass-action chemical reaction networks at steady state) and describe a framework for their analysis based on optimization using numerical algebraic geometry. Specifically, we use probability-one polynomial homotopy continuation methods to compute all critical points of the objective function, then filter to recover the global optima. Our approach exploits the geometric structures relating models and data, and we demonstrate its utility on examples from cell signaling, synthetic biology, and epidemiology.

\end{abstract}

\section{Introduction}
Across the physical, biological, and social sciences, mathematical models are formulated and studied to better understand real-world phenomena. Often, multiple models are developed to explore alternate hypotheses.  It then becomes necessary to choose between different models, for example, based on their fit with noisy experimental data. This is the problem of model selection, a fundamental scientific problem with practical implications \cite{chamberlin:science:1965,Burnham:Springer:2002,Kirk:2013hq}.


The standard approach to model selection is to first estimate all model parameters and hidden variables from the data, then select the model with the smallest best-fit error, up to some penalty on model complexity \cite{Liepe:2014iw,Lillacci:2010jm}. Thus, at its core, model selection is intimately tied to parameter estimation, which, for a given model $f(a, x) = 0$ in the parameters $a$ and variables $x$ with measurable ``outputs'' $y = g(x)$, say, can be written as the following optimization problem:
\begin{align}
 \min_{a, x, y} \| y - {y}' \|^{2} \quad \text{s.t.} \quad
 \left\{
 \begin{array}{r@{\ }c@{\ }l}
  f(a, x) & = & 0\\
  y & = & g(x)
 \end{array} \right.
 \label{eqn:opt-general}
\end{align}
where $y'$ denotes the observed data, i.e. measured outputs. Unless $f$ and $g$ are convex, solving \eqref{eqn:opt-general} is a nonconvex problem, which can be challenging as standard local solvers run the risk of getting trapped in local minima (especially in high dimensions). This can be mitigated somewhat with techniques such as simulated annealing \cite{kirkpatrick:1983:science,cerny:1985:j-optim-theory-app} or convex relaxation which has been successful for model invalidation \cite{rumschinski2010set,Melykuti:2010eu,anderson:2009}, but there is generally no guarantee that a global minimum will be found.


When $f$ and $g$ are polynomial, however, problem \eqref{eqn:opt-general} can be solved globally by finding all roots of an associated polynomial system. In this case, ideas from computational algebra and algebraic geometry can be effective; see, e.g., \cite{Manrai:2008kb,Ho:2010hb,Harrington:2012us,MacLean:2014ar} for applications of Gr\"{o}bner bases in systems biology and \cite{DSS09} for applications of algebraic geometry to statistical inference. Such symbolic methods tend to be computationally expensive, which limits their use in practice. Thus, although they provide a solution in principle, new algorithms and techniques are yet desired.


In this paper, we aim to fill this gap by proposing a framework for global parameter estimation for polynomial deterministic models using numerical algebraic geometry (NAG), a suite of tools for numerically approximating the solution sets of multivariate polynomial systems via adaptive multi-precision, probability-one polynomial homotopy continuation \cite{SW,bertini}. Our approach  scales well in dimension relative to classical symbolic methods~\cite{BertSing}, and, while it comes with a higher computational cost than standard local optimization, it has a probability-one guarantee to recover the global optima, solving problem \eqref{eqn:opt-general} in the strong sense.  
This allows us to reason rigorously about model fit and to address the related problems of model selection and parameter estimation from a maximum-likelihood perspective.


We demonstrate our techniques on examples from biology, where polynomial models often arise as the steady-state descriptions of mass-action chemical reaction networks. 
Although some limitations remain, we believe that this work achieves its primary purpose of introducing NAG 
as a valuable complement
to existing tools for model evaluation and analysis.  Additionally, this paper highlights specific challenges that arise when using polynomial methods for model inference, such as dealing with positivity constraints and non-isolated solutions, and provides guidance for tackling these challenges.

The remainder of the paper is organized as follows. In the next section, we state precisely the problems with which we are concerned: model validation, model selection, and parameter and hidden variable estimation. We then present NAG algorithms for solving each problem. Finally, we showcase our approach on a few examples including cell death activation, synthetic bio-circuits, HIV progression, and protein modification.

\section{Problem Statement}

Consider a model whose dynamics are described by the system of polynomial differential equations
\begin{align}
\dot { x} & = f({a}, {x})
\label{eq:dynamics1}
\end{align}
where ${a}=(a_1, \ldots, a_k)$ are parameters (e.g. rate constants in a deterministic mechanistic model, such as a chemical reaction network with mass-action kinetics), ${x}=(x_1, \ldots, x_n)$ are variables, and $f=(f_1, \ldots, f_r)$ are polynomials in ${x}$ and ${a}$ with measurable outputs $y=g(x)$ where $y=(y_1, \ldots, y_m)$, $m \leq n$, and $g=(g_1, \ldots, g_m)$ are polynomials in $x$.  While the parameters $a_1, \ldots, a_k$ are treated as fixed variables in our exposition, we separate them from $x_1, \ldots, x_n$ to respect how such variables are treated differently in experimental and computational settings.

In algebraic geometry a \emph{variety} is a solution set of a system of polynomial equations; we use this terminology for our next two definitions. The \emph{real model variety} is the solution set of the system
\begin{align}
f &= 0\\
y-g(x) &=0, \quad \text{that is,}
\end{align}
 $$(\mathcal V_{\mathcal M})_{\mathbb R} := \{ (a, x, y) \in \mathbb R^{k+n+m} \ : \ f(a, x)=0, \ y-g(x)=0\}$$
corresponding to the steady states of the model.  Now, consider, for simplicity, the case of a single data point $ \hat{y}=( \hat y_1, \ldots, \hat y_m)$. (See SI Appendix for multiple data points.) The \emph{real data variety} is then the  affine linear space
$$(\mathcal V_{\cD})_{\mathbb R} := \{ ( a, x, y) \in \mathbb R^{k+n+m} \ :  y_i=\hat y_i, \ \forall i=1,\ldots,m \},$$
with $\dim (\mathcal V_{\cD})_{\mathbb R} = k+n$. We consider the case when the data includes some extrinsic (measurement) noise; we assume the errors $\{ \epsilon_1, \ldots, \epsilon_m \}$ on the observed data variables are uncorrelated random variables and each error $\epsilon_i$ is normally distributed with known variance $\sigma_i$ (which can be obtained by instrument calibration).

Using this geometric framework, the problems of (1)~model validation, (2)~model selection, and (3)~parameter estimation, can be described precisely in terms of the real algebraic varieties $(\mathcal V_{\mathcal M})_{\mathbb R}$ and $(\mathcal V_{\cD})_{\mathbb R}$.

\subsection{Problem 1: Model Validation}

For model validation, we want to determine whether a deterministic polynomial model $\mathcal M$  is  compatible with the data according to a given significance level~$\alpha$. This is akin to asking whether the model is a ``good fit" for the data.  A natural goodness-of-fit statistic is
:
\begin{align}
 d^2:=\min & \sum_{i=1}^m \frac{(y_i - \hat y_i)^2}{\sigma_i^2} \label{eq:modelCompMin}
 \text{ subject to }  (a, x, y) \in (\mathcal V_{\mathcal M})_{\mathbb R}.
\end{align}

When the variances differ, $d^2$ is the minimum squared weighted Euclidean distance between $(\mathcal V_{\mathcal M})_{\mathbb R}$ and $(\mathcal V_{\cD})_{\mathbb R}$.  When all variances are equal to one, the value of \eqref{eq:modelCompMin} is just the minimum squared distance.  For the remainder of the text, we will assume the latter, knowing that we can simply rescale.

Optimization problem~\ref{eq:modelCompMin} can be derived directly from the log-likelihood function
\begin{align*}
 \log L(a, x, y) = \sum_{i = 1}^{m} \left( \frac{1}{2} \log 2 \pi \sigma_{i}^{2} - \frac{(\hat{y}_{i} - y_{i})^{2}}{2 \sigma_{i}^{2}} \right)
\end{align*}
as demonstrated in Section 2.1 of the SI.  This provides a bridge to standard statistical model selection tools such as AIC~\cite{AIC} and BIC~\cite{BIC}.

If the data are generated from the model $\mathcal M$ with normally distributed extrinsic noise, the distribution function of $d^2$ is dominated by that of the chi-squared distribution with $m$ degrees of freedom, $\chi^2_m$, where $m$ is the number of measurable outputs. Thus, if $p_{\alpha}$ is the upper $\alpha$-percentile for $\chi_m^2$, then $\Pr(d^2 \geq p_{\alpha}) \leq \Pr(U \geq p_{\alpha})=\alpha$ where $U \sim \chi^2_m$.  We reject the model $\mathcal M$ as \emph{incompatible} if $d^2 > p_{\alpha}$; otherwise we say that the model $\mathcal M$ is {\em compatible}.

If the real model and data varieties intersect, that is, $(\mathcal V_{\mathcal M} )_{\R}~\cap~(\mathcal V_{\mathcal D})_{\R} \neq \emptyset$, then $d^2=0$, and we also say that the model is compatible with the data.  If there are restrictions on $({ a}, x, y)$, for example if all parameters and variables are required to be non-negative, then finding $d^2$ becomes a \emph{constrained optimization} problem (see SI Appendix).

\subsection{Problem 2: Model Selection}

For model selection, we are given a set of models $\{ \cM_1, \ldots, \cM_s\}$ and want to determine the model of best fit. In this setting, we use the statistic $d^2$ to make a selection, either by choosing the model with the smallest goodness-of-fit statistic or by using $d^2$ in conjunction with a complexity penalty, similar to the Bayesian or Akaike information criteria \cite{Burnham:Springer:2002}.

If the statistic $d^2$ evaluates to zero for all (or even multiple) models, then we are unable to make a selection between the models.  This can be remedied by designing experiments that yield more informative data.  For example, measuring more variables can reduce the dimension of $(\V_{\mathcal M} )_{\R} \cap (\V_{\mathcal D})_{\R}$; the most informative situation is when $(\V_{\mathcal M_i} )_{\R} \cap (\V_{\mathcal D})_{\R} = \emptyset$ for all models. This indeterminacy can also be resolved by taking multiple measurements and minimizing the joint squared distance (see SI Appendix).

\subsection{Problem 3: Parameter Estimation}

  Parameter estimation can be achieved by finding the point $( a^*, x^*, y^*) \in (\V_{\mathcal M})_{\mathbb R}$  that minimizes the value of \eqref{eq:modelCompMin}.  The point $( a^*, x^*, y^*)$ is the maximum likelihood estimate under the given noise assumptions. The parameter estimate is then $a^*$; the estimate of the hidden variables, $x^*$; and the estimate of the ``de-noised" outputs, $y^*$.  Of course, if the data and model varieties intersect, there will be one or more (possibly infinite) choices for $(a^*, x^*, y^*)$.  Otherwise, it is a matter of solving a polynomial system that yields the point(s) on
$(\V_{\mathcal M})_{\mathbb R}$ nearest $(\V_{\mathcal D})_{\mathbb R}$.  This is
described in more detail in the next section.



\section{Geometry}

In each of the problems stated above, we seek to minimize the distance between $(V_{\mathcal M})_{\mathbb R}$ and $(V_{\cD})_{\mathbb R}$ or to
find the intersection of these two sets.
Standard methods for solving non-linear optimization problems are local in nature, i.e., only guaranteed to converge to a local minimum which may or may not be the global minimum.
However, using NAG, we find {\em all} local extrema over $\mathbb C$, necessarily including the global minimum.  Due to this global benefit, NAG has been used before in statistical inference in the field of algebraic statistics \cite{GPV:2013, Hauenstein:2014}.


 Let $V_{\mathcal M} \subseteq \C^{k+n+m}$ be the (complex) Zariski closure of $(V_{\mathcal M})_{\mathbb R}$ and $V_{\cD} \subseteq \C^{k+n+m}$ be the (complex) Zariski closure of $(V_{\cD})_{\mathbb R}$.  We will refer to $V_{\cM}$ and $V_{\cD}$ as the \emph{model variety} and \emph{data variety}, respectively.  

 The problem of determining the intersection of $V_{\mathcal M}\cap V_{\cD}$ is simply a matter of solving the polynomial system obtained by taking the union of the polynomials defining $V_{\mathcal M}$ and the polynomials defining $V_{\cD}$.  This is  handled directly by NAG.  
 If the intersection is nonempty and positive-dimensional (complex curves, surfaces, etc), real points can be found using the polynomial homotopy method described in \cite{hauenstReal}, a method based on symbolic algorithms in \cite{Rouillier:2000}, and, more classically, on the decision method  in \cite{Seidenberg:1954}.  


The problem of finding the points on the two varieties nearest one another can also be stated in terms of a polynomial system, on which we then call NAG solvers to find solutions.  A well-known necessary condition for local extrema is given by Lagrange multipliers.  In the main text we assume that $r+m= \text{ codim } \V_{\mathcal M}$; however, when this is not the case, the number of equations can be reduced (see SI Appendix). 

\begin{proposition}[Equations given by Lagrange multipliers]  Let $c=$ codim $V_{\cM}$ and let $f'=(f'_1, \ldots, f'_c)=0$  on a Zariski open set of $V_{\mathcal M}$.  If $(a, x, y) \in (\V_{\mathcal M})_{\R}$ is a local minimum of
\begin{equation}\label{eq:weighteddistance}
 \sum_{i=1}^m (y_i - \hat y_i)^2,
 \end{equation}
then there exists ${ \lambda}:=(\lambda_1, \ldots, \lambda_c)$, such that $(a, x, y, \lambda)$ is a solution to the system
\begin{align}
&f'=0, \label{eq:system1}\\
& y - g(x)=0 \label{eq:out}\\
&\lambda_1 \nabla f'_1 + \ldots \lambda_c \nabla f'_c +  \left[\begin{array}{c}0 \\y - \hat y\end{array}\right] =0 \label{eq:system2}.
\end{align}
\end{proposition}



Solving this system with NAG will provide us with all local extrema of \eqref{eq:weighteddistance} over $\V_{\mathcal M}$,
 from which we may easily select the pair of nearest points. The geometry of zero sets of systems such as~\eqref{eq:system1}-\eqref{eq:system2} are described in \cite{Draisma:2015}.

\subsection{Numerical Algebraic Geometry}

Given a polynomial system $F$ consisting of $r$ polynomials and $N$ variables,  NAG packages, such as {\tt Bertini} \cite{bertini}, {\tt PHCpack} \cite{PHCpack}, {\tt HOM4PS-2.0} \cite{HOM4PS}, use polynomial homotopy continuation to provide {\em probability-one} numerical methods for finding
approximations of all isolated complex solutions of $F=0$ (points) as well as {\em witness points} on all positive-dimensional irreducible components of the solution set of
$F=0$.  These methods are probability-one in that the computations include some randomization, and this randomization will yield theoretically correct results so long
as the random numbers are not chosen from some measure zero set in the parameter spaces of potential choices \cite{SW,SIAMbook}.

If $x\in\mathbb R^N$ is a {\em real} solution of $F=0$, it is either isolated among the complex solutions or it lies on a positive-dimensional complex irreducible component.  In the
former case, the methods of NAG will find $x$ and recognize it as real.  In the latter case, $x$ can be difficult to uncover.
However, for our purposes, we usually only need to verify the existence of a real solution.  In this case, we can find witness points on all positive dimensional components and then use the procedure in Section 2.1 of \cite{hauenstReal} to verify the existence of real points.



\subsection{Algorithms}

We present three related algorithms to address model validation, model selection, and parameter estimation (see Fig.\ 1). 

The aim of the first algorithm, Algorithm 1, is to find the pair of points that minimize the distance between $(\mathcal{V_M})_{\mathbb{R}}$ and $(\mathcal{V_D})_{\mathbb{R}}$. If $(\mathcal{V_M})_{\mathbb R} \cap (\mathcal{V_D})_{\mathbb R} = \emptyset$, then this is obtained by solving \eqref{eq:system1}-\eqref{eq:system2}; otherwise, additional techniques are required.

We note also simply that these techniques are indeed probability-one algorithms.  The computations will necessarily be carried out in finite time and the steps proceed linearly (no loops), so the methods will necessarily finish.

\subsubsection{Algorithm 1:  Model validation}
\begin{tabularx}{\linewidth}%
 {l>{\setlength\hsize{.99\hsize}\raggedright}X%
   >{\setlength\hsize{.01\hsize}\raggedright}X}
\hline
{\bf Input}&Model $\cM$, data $\cD=\{\hat y\}$, tolerance $\alpha$&\\
{\bf Output}&{\em yes} or {\em no}&\\
\hline
1 & If $\mathcal{V_M}\cap\mathcal{V_D}=\emptyset$, goto Step 3. &\\
2 & If $\dim(\mathcal{V_M}\cap\mathcal{V_D}) \geq 0$ and &\\
   &$(\mathcal{V_M})_{\mathbb R}\cap(\mathcal{V_D})_{\mathbb R} \neq \emptyset$, return         \emph{yes}, &\\
   & else, goto Step 3. &\\

3 & Find a pair $(z_1,z_2)\in (\mathcal{V_M})_{\mathbb R}\times(\mathcal{V_D})_{\mathbb R}$ that minimizes \eqref{eq:weighteddistance} (using NAG software such as {\tt Bertini} or {\tt PHCpack}). &\\
4 & If $||z_1-z_2||^2 < p_\alpha$, return {\em yes}; else {\em no}.&\\
\hline\\
\end{tabularx}


The computation of the intersection $\mathcal V_{\mathcal M} \cap \mathcal V_{\mathcal D}$ in Steps 1 and 2 can be determined in several ways.  The simplest way is by considering dimensions: if $\dim(\mathcal{V_M})+\dim(\mathcal{V_D})$ exceeds the ambient dimension, they will almost surely intersect.  If the ambient dimension is larger than the sum of the variety dimensions, they typically do not intersect.  To compute the intersection (or check to see if it is nonempty), one could substitute $\hat y - g(x)$ for \eqref{eq:out} in the system \eqref{eq:system1}-\eqref{eq:out}.

In Step 2, if $\dim(\mathcal{V_M}\cap\mathcal{V_D})=0$, then the intersection of the two varieties consists of finitely many points; the condition $(\mathcal{V_M})_{\mathbb R}\cap(\mathcal{V_D})_{\mathbb R}\neq\emptyset$  indicates that at least one of the points is real, which is straightforward to determine.  If $\dim(\mathcal{V_M}\cap\mathcal{V_D})>0$,  to check if $(\mathcal{V_M})_{\mathbb R}\cap(\mathcal{V_D})_{\mathbb R}\neq\emptyset$, one needs more sophisticated methods, such as those in  \cite{hauenstReal}.

To find the pair $(z_1,z_2)$ in Step 3, one may solve the polynomial system \eqref{eq:system1}-\eqref{eq:system2}.  
If there is a positive-dimensional set of (complex) extremal points, then the procedures in~\cite{hauenstReal} could be used to determine if the set contains a real point. 

If there are constraints on the variables or parameters, for example, if we seek to minimize \eqref{eq:weighteddistance} over the non-negative part of $(\mathcal V_{\mathcal M})_{\mathbb R}$, then the algorithm is updated as follows:  if the $\dim( \mathcal{V_M} \cap \mathcal{V_D}) = 0$ or $\mathcal{V_M} \cap\mathcal{V_D} = \emptyset$ then the Karush-Kuhn-Tucker equations as described in Section 3.1 of the SI Appendix should be used, while if $\dim( \mathcal{V_M} \cap \mathcal{V_D}) >0$, or if there is a positive-dimensional set of extremal points at Step 3, then the algorithm should return \emph{possibly}. There are methods~\cite{LBSW,BDHSW,BertiniReal} for finding real points, curves, and surfaces within complex components of dimension 2 or less, but little is known about higher dimensions.




\subsubsection{Algorithm 2:  Model selection}
In this case, there are several competing models, each with its own polynomial system.
The algorithm proceeds much as in Algorithm 1, but iterated for the several models under consideration.  If a threshold $\alpha$ is provided, one should first reject models that do
not adequately support the data (
$d^2 \geq p_\alpha$), then choose the model yielding the minimum value of $d^2$ (up to some complexity penalty).  Various conclusions may be drawn, e.g., {\em no model adequately fits the data} or {\em three models adequately fit the data and model $\mathcal M_1$ provides the best fit}. 

\subsubsection{Algorithm 3:  Parameter estimation}
Again, this algorithm is similar to the first.  The input consists of only one model $\cM$ and data $\cD$.  It is assumed that there are unknown parameters and the goal is to
find values of these parameters producing the best fit between $\cM$ and $\cD$.  The output of Steps 2 and 3 need to be adjusted appropriately. The output of Step 4 is simply
$z_2$ since there is no $\alpha$ to be used for rejection.
The method also simultaneously estimates hidden/unknown variables and ``de-noised'' outputs.


\subsubsection{Simple Example}
To illustrate Algorithm 1, consider a simple model with three variables $x,y,z$ and three parameters $a,b,c$ satisfying $$\frac{x^2}{a^2}+\frac{y^2}{b^2}+\frac{z^2}{c^2} = 1.$$ Let $\alpha=0.1$ and assume, for this example, that the variances on the errors are $\sigma_i^2=0.1$ for $i=1,2,3$. The model variety $\mathcal{V_M}$ is simply an ellipsoid (Fig.\ 2A) where $a,b,c$ describe the principal axes. Suppose we know that $a,b,c=1$. For the case when the outputs are $x$ and $y$ and $\hat x=0$, $\hat y = 0$ and $\alpha=0.1$, Step 2  indicates that the data {\em does} fit the model, i.e. the model is compatible with the data.  In this case, there are two real points in the 0-dimensional intersection $\mathcal{V_M}\cap\mathcal{V_D}$ (see Fig.\ 2B). For the same set-up, but with data $\hat { x}=0$, Step 3 indicates that this data {\em possibly} fits this model.  Since there is a positive-dimensional intersection (Fig.\ 2C), it is possibly compatible (depending on constraints imposed by the user). For the same model and $\alpha$, but different data $\hat x=1.7, \hat y=0$, Step 4 yields points $(1,0,0)$ and $(1.7,0,0)$, so $||z_1-z_2||^2>0.4605=p_\alpha$, and the model is rejected (Fig.\ 2D). However when the data are $\hat x=1.01,\hat y=0$, Algorithm 1 indicates model compatibility (Fig.\ 2E). Previous algebraic methods that required Gr\"{o}bner basis calculations would result in a zero set and thus those approaches are not useful here.


\section{Results}
We demonstrate our methods on problems in biomedicine: cell death activation, synthetic biology, epidemiology and multisite phosphorylation. Each of the forthcoming applications can be written as a mass-action chemical reaction network, which has the form $\dot{x} = f(a, x)$, studied at steady state: $f(a, x) = 0$ as in the problem statement.
Throughout these real-world examples, we emphasize the pivotal computations in these methods, such as determining the dimension of the intersection $V_{\mathcal M}\cap V_{\cD}$ and finding points in the intersection (see Fig.\ 1B). Our first two examples, cell death signaling and genetic toggle bio-circuits, demonstrate how to handle positive dimensional intersections using two different approaches, while the remaining two examples, HIV and MAP kinase signaling, highlight analysis of zero-dimensional and empty intersections. In the following examples, we are interested in results that can be interpreted biologically, therefore we restrict our analysis to  non-negative real solutions.

\subsection{Cell death activation}

We demonstrate model compatibility (Algorithm 1) on an example from receptor-mediated programmed cell death, which is initiated by the activation of death receptors upon the detection of extracellular death ligands \cite{Thomson:1995,Raff:1998,meier:n:2000,Fulda:2006}. We consider, in particular, the ``cluster'' model of \cite{Ho:2010hb}, which was inspired by crystallographic data \cite{scott:nat:2009} and describes the recruitment of receptors by ligands into local self-activating clusters capable of bistability. 


The cluster model is a system of $3$ degree-$4$ polynomials in the form of \eqref{eq:dynamics1} in $3$ variables (representing various receptor states) and $6$ rate parameters, supplemented by ligand and receptor conservations (see SI Appendix). We assume that we can measure the total ligand and receptor concentrations, which may be considered experimental inputs, as well as the concentration of active receptors. We do not assume access to the concentrations of other individual receptor states nor to any of the rate parameters.

A steady-state data point was simulated from the model with all parameters and initial concentrations drawn independently and identically from the log-normal distribution $\ln \mathcal{N} (0, 4)$, then combined and corrupted with i.i.d.\ noise from $\mathcal{N} (0, 0.1)$ to obtain $\hat{y}$. The real model and real data varieties intersect in the positive orthant with a distance zero and hence the model is indeed compatible with the data.




\subsection{Synthetic biology and experimental design}
We demonstrate an example from synthetic biology with excess intersection ($\dim(\mathcal{V_M}\cap\mathcal{V_D})>0$). A goal in synthetic biology is to design or modify existing biological systems with new features according to specific design criteria. Reverse engineering of biological systems often includes modules (such as feedback loops) and how these are interconnected are described by different circuits (models). Understanding differences between bio-circuit implementations is crucial; therefore we compare three bistable bio-circuits models analyzed in \cite{Siegal:arxiv:2014}:  monomer-dimer toggle circuit ($\mathcal M_1$), dimer-dimer  toggle circuit ($\mathcal M_2$), and single operator gene circuit ($\mathcal M_3$), which were initially presented in \cite{Gardner:2000:nature,Hasty:chaos:2001,Siegal:PLoSCB:2011}. The model variables include genes ($X_i$) and proteins ($P_i$) where $i = 1$ for $\mathcal M_3$ or $i = 1,2$ for $\mathcal M_1, \mathcal M_2,$ as well as these species complexes (e.g., $P_iP_i, X_jP_iP_i$). These variables interact following mass-action kinetics and form systems of polynomial differential equations where $\mathcal M_1, \mathcal M_2,$ and $\mathcal M_3$ have 7, 8 and 6 model variables, respectively, and 10, 12, and 9 kinetic parameters, respectively. The models can be reduced (given in SI Appendix) by assuming that the total amount of gene 1 ($X_{1_{tot}}$) and gene 2 ($X_{2_{tot}}$) is conserved.

Suppose that the total amounts $X_{1_{tot}}$ and $X_{2_{tot}}$ and specific protein synthesis and degradation parameters $k_{bas_1}$, $k_{bas_2}$, $k_{deg_1}$, and $k_{deg_2}$ are known. Since protein concentrations are often measurable, we assume that our data are $P_1$, $P_2$, and their complexes $P_1P_1$, and $P_2P_2$. The aim is to select the best model $\mathcal M_1$, $\mathcal M_2$, and $\mathcal M_3$ given the data.  We simulate steady-state data from the dimer-dimer toggle model ($\mathcal M_2$) and add Gaussian noise from $\mathcal{N}(0,0.1)$. 
We find that all three have positive dimensional intersections, where the dimension of the intersections are:
\begin{align*}
\dim((\mathcal{V_{M_\text{1}}})_{\mathbb R}\cap(\mathcal{V_{D}})_{\mathbb R})&=3, \\
\dim( (\mathcal{V_{M_\text{2}}})_{\mathbb R}\cap(\mathcal{V_{D}})_{\mathbb R})&=4, \\
\dim( (\mathcal{V_{M_\text{3}}})_{\mathbb R}\cap(\mathcal{V_{D}})_{\mathbb R})&=3.
\end{align*}
Clearly, all three models are compatible with the data, thus one can only select a ``best fit" model using data-independent measures, e.g. number of parameters, dimension, etc.


In fact, the dimensions of the model and data varieties can help us design more informative experiments for model selection.  For example, the variety associated to the monomer-dimer toggle model lives in a 15 dimensional ambient space, $\V _{\mathcal M_1} \subseteq \C^{15}$, and has dimension 6, while the data variety $\V _{\mathcal D} \subseteq \C^{15}$ has dimension 12; thus by the Dimension Theorem, we can determine that $\text{dim} (\V _{\mathcal M_1} \cap \V _{\mathcal D}) \geq 6 + 12 - 15=3$.  This dimension calculation provides guidance towards the number of minimal additional variable and parameter values that must be measured to ensure $\mathcal{V_{M}}_1\cap\mathcal{V_D}=\emptyset$, i.e., at least four more variables and parameter values must be known.

Suppose we can experimentally measure four forward biochemical reaction rate constants (e.g., $k_{cF}$, $k_{kF}$, $k_{nF}$, and $k_{kR}$), then $\V _{\mathcal M_1}$ is cut down by four dimensions and does not intersect $\V_{ \mathcal D}$. We get similar results for the model varieties associated to $\mathcal M_2$ and $\mathcal M_3$ provided we measure rate constants specific to these models (see SI Appendix).  Now that all the intersections are empty, we run Algorithm 2 and find that the sum of squares (Eq.~\eqref{eq:modelCompMin}) for each model are as follows:  $d_1^2= 0.2262$, $d_2^2=7.34 \times 10^{-7}$, and $d_3^2=0.3040$.  Therefore, we select the $\mathcal M_2$ model, which is indeed the true model.

\subsection{HIV Progression}
We demonstrate parameter estimation (Algorithm 3) on an example coming from epidemiology. We use a model that includes long-term HIV dynamics from initial viremia, latency, and virus increase \cite{Mehta}, based on \cite{HCV07}. In the model (see SI Appendix), the HIV virus inhibits the CD4+T cell population while promoting macrophage proliferation, which in turn houses the replicating virus. As macrophages proliferate, the virus reservoir increases, so the model offers a description of HIV patient progression to AIDS. Model variables ${x}$ are uninfected CD4+T cells ($T$), infected CD4$^+$ T cells ($T_i$), uninfected macrophages ($M$), infected macrophages ($M_i$), and HIV virus population ($V$).

Hernandez-Vargas et al.\ show that the model can have two real equilibria, one of which is stable, representing patients that are ``long-term non-progressors" \cite{Mehta}.  The parameters ${a}$ are $(s_1, s_2, k_1, \ldots, k_6, \delta_1, \ldots, \delta_5)$, where $s_i$ represents synthesis of T cells and macrophages, $k$ are rate constants describing interactions between variables $x$, and $\delta_i$ represents natural death. For this example, $y=x.$
 We estimated the natural death of the virus, parameter $\delta_5$, using the long-term non-progressors steady-state value (Table 3 of \cite{Mehta}) and adding noise to each variable $\sim \mathcal{N}(0,1)$. By Algorithm 3, the data variety and model variety do not intersect. We find the closest point and estimate $\delta_5^*=2.99876$ (true value of $\delta_5 = 3$).


\subsection{Multisite phosphorylation with experimental data}
We examine phosphorylation mechanisms of cellular signaling with experimental data, and demonstrate model selection (Algorithm 2) and parameter estimation (Algorithm 3). We focus on phosphorylation, a key cellular regulatory mechanism that has been the subject of extensive study, both experimentally and theoretically (\cite{futran:2013} and references therein). An area of interest is the mechanism by which a kinase phosphorylates a two-site substrate, either distributively, where the kinase can add at most one phosphate before dissociating, or processively, where it can add both phosphates in sequence. The MAPK/ERK pathway is a well-known system for studying phosphorylation, whereby MEK (kinase) phosphorylates ERK (its substrate). Aoki et al.\ \cite{Aoki:2011ji} showed experimental evidence, while working with polynomial models, that the mammalian MAPK/ERK pathway acts distributively {\it in vitro} but processively {\it in vivo}.

We compare these distributive and processive models
against the {\it in vivo} data reported in the same study. The distributive model consists of 12 molecular species and $17$ mass-action reactions, while the processive model has 14 species and $18$ reactions (species correspond to variables, each reaction corresponds to a parameter). The data take the form of $36$ concentration measurements of three aggregate phosphoforms over a range of 12 EGF stimulation levels. All model parameters are calibrated using {\it in vitro} estimates by \cite{Aoki:2011ji}, except the parameter $k_{1}$ representing EGF loading, which we estimate for both models (see SI Appendix).

Next we perform model selection by running Algorithm 2 on each data point individually and select independently for each run the preferred model (more details in SI Appendix). Under low EGF stimulation, the best model estimates are nearly identical with a slight preference for distributive. At high EGF stimulation, the models are identical with no preference for one model over the other.  These results can be justified by noting that the main distinction between distributivity and processivity is nonlinear switching behavior (i.e., a sigmoidal response curve), and this only occurs at intermediate stimulations.
However, at medium EGF stimulation (see Fig 3), there is a preference for the processive model, which supports the findings in \cite{Aoki:2011ji}.

\section{Conclusion}

The problem of determining whether given real-world data fits one or more given mathematical models is challenging.  When a model is defined by
algebraic (polynomial) functions, the methods of NAG may be employed to study the geometry underlying the model and data.
In particular, these methods are useful for model variables observed at steady-state. We demonstrated this numerical and geometric framework for comparing models with experimental dose-response data in MAPK/ERK pathway and highlighted that the intermediate EGF doses were the most informative for model selection, complementing another finding that model selection results can be very sensitive to experimental parameters \cite{Silk:2014}. 

Despite the difficulties associated with positive dimensional components, and limitations in analysis, we can reproduce compatible models,
and furthermore can predict additional information, such as measurements of parameters, that are necessary for selecting models. Our geometric investigations of positive dimensional components may perhaps relate to algebraic analyses for biochemical models, such as model identifiability or matroids for experimental design \cite{Gross:2015vo,MacLean:2014ar,MeshkatS14}. 

There are further directions to be considered in this vein, aside from making the existing computational methods more efficient.  First, there would be great value in
developing strictly {\em real} geometric methods for solving polynomial systems such as those that appear in this article.  Some such techniques exist, but only in very
special circumstances.  Second, there would be much value in developing effective numerical methods for treating inequalities.  It should be noted that the methods
described in~\cite{BPR} and the references therein will incorporate such constraints, though the cost of such computations restricts their use to relatively low dimensions.
Finally, there is likely much to be gained from considering the geometry underlying models not defined by algebraic functions.  Algebraic geometry provides very clean,
well-understood structures, paving the way for numerical methods.  Differential geometry or topology could lead to similarly useful techniques for model selection and parameter estimation.



\section{Materials and Methods}

\subsection{Numerical algebraic geometry}
General references for NAG include~\cite{SW,SIAMbook}, with the
latter doubling as a user manual for the software package Bertini. For computations, we used Bertini 1.4 and Macaulay2 version 1.6.

\subsection{Data generation}
Data simulated from cell death activation, synthetic biology and HIV models were performed in MATLAB R2014b using ode15s.


\section{Acknowledgements}
We thank Kazuhiro Aoki for discussions and providing model and data files for the MAP Kinase example.
EG, KLH, and HAH acknowledge funding from the American Institute of Mathematics (AIM).  EG was supported by the US National Science Foundation grant DMS-1304167. BD was partially supported by NSF DMS--1115668. KLH acknowledges support from NSF DMS-1203554.  DJB gratefully acknowledges partial support from NSF DMS--1115668 and the Mathematical Biosciences Institute (MBI).   HAH gratefully acknowledges funding from AMS Simons Travel Grant, EPSRC Fellowship EP/K041096/1, King Abdullah University of Science and Technology (KAUST) KUK-C1-013-04 and MPH Stumpf Leverhulme Trust Grant.


\clearpage

\begin{figure}[h!]
\includegraphics{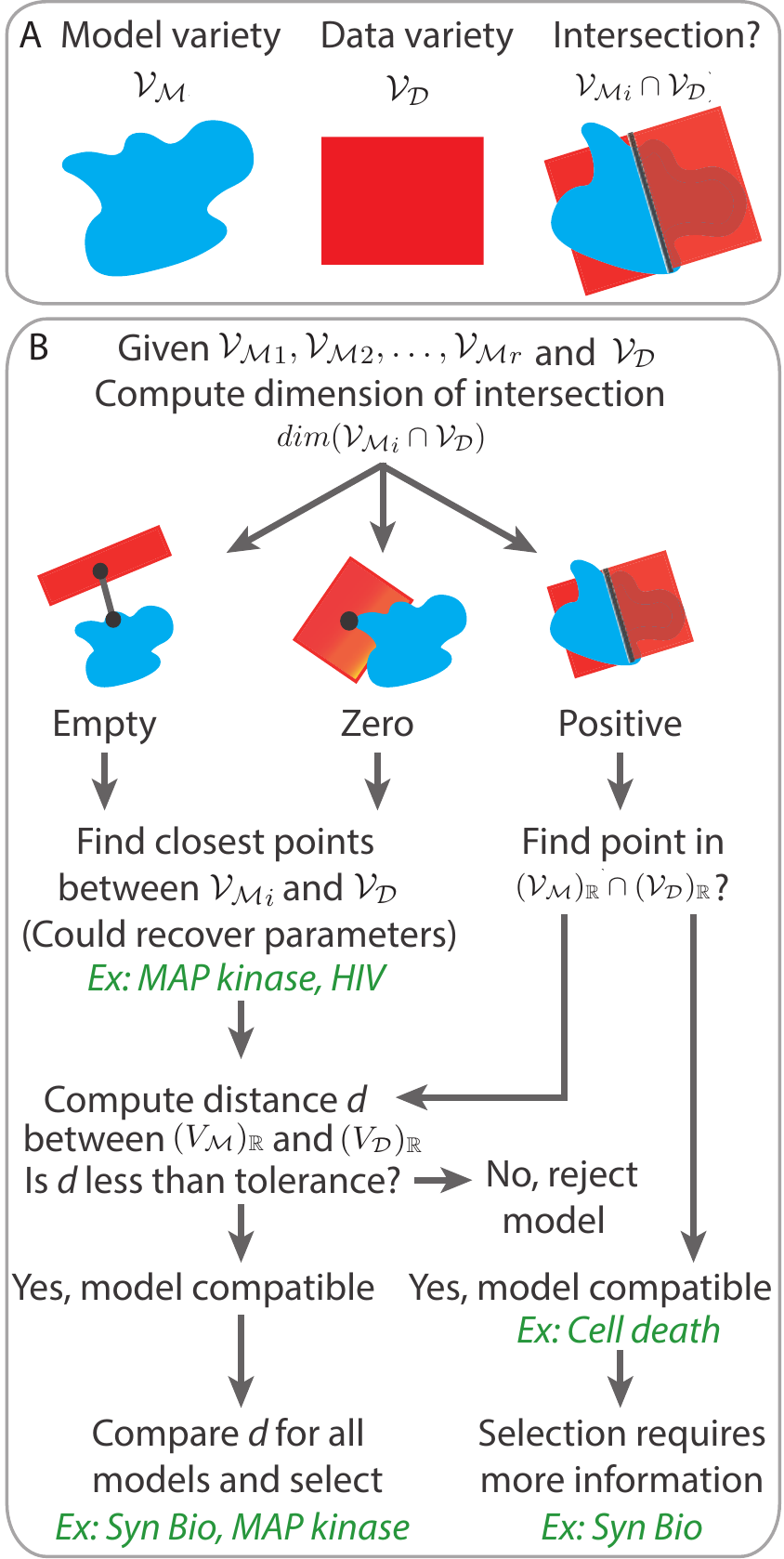}
\caption{Schematic of numerical algebraic geometry framework corresponding to Algorithms I--III. (A) Input to algorithms include model (system of polynomials) translated into a model variety (red), and steady-state data translated into a data variety (blue). (B) Flow chart of model compatibility, parameter estimation and model selection methods. Examples (green) are described in Results section.}
\label{fig-traj}
\end{figure}

\begin{figure}[h!]
\includegraphics{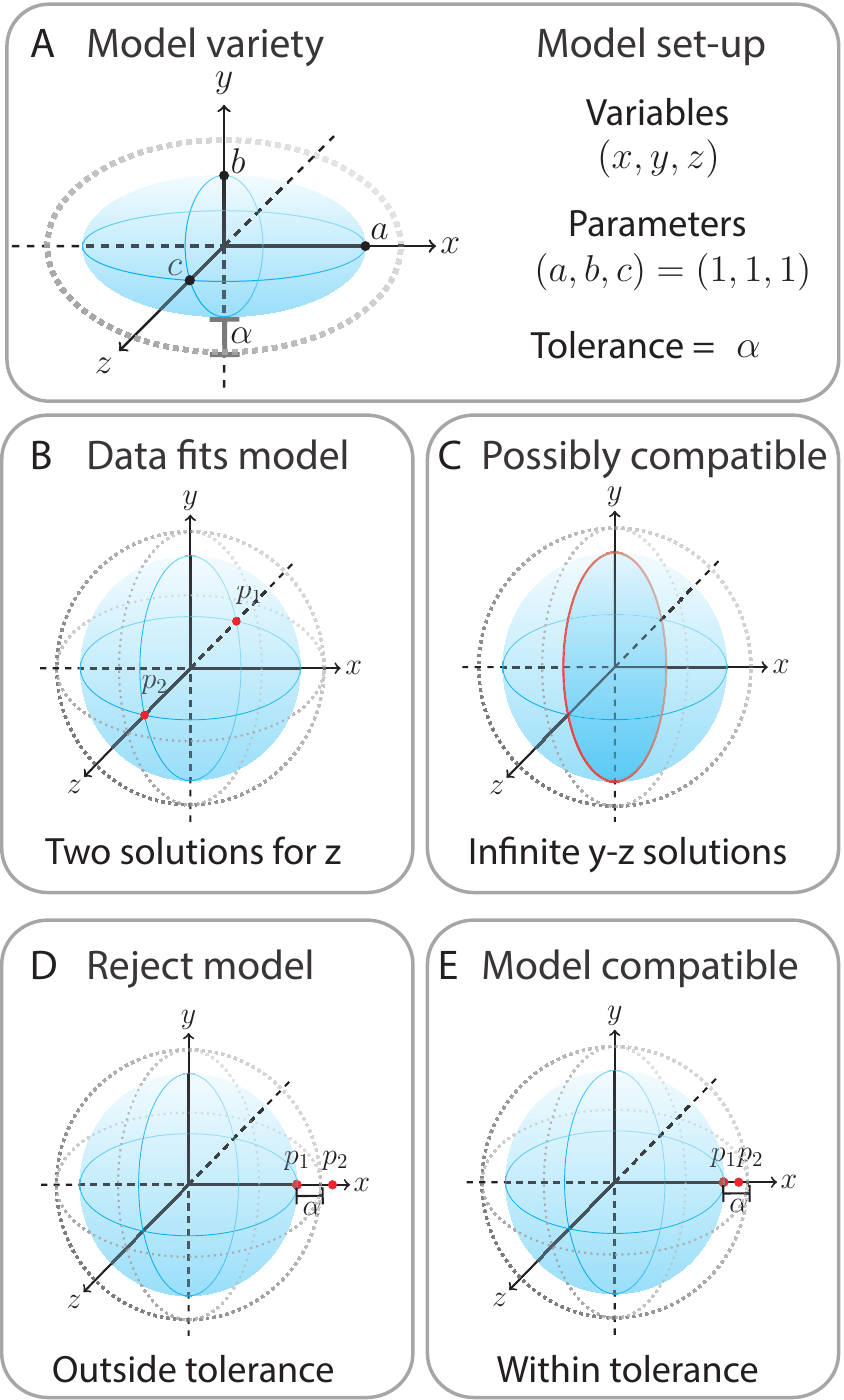}
\caption{Simple example demonstrating model compatibility following Algorithm 1.}
\label{fig-compat}
\end{figure}

\begin{figure}[h!]
\includegraphics[scale=0.6]{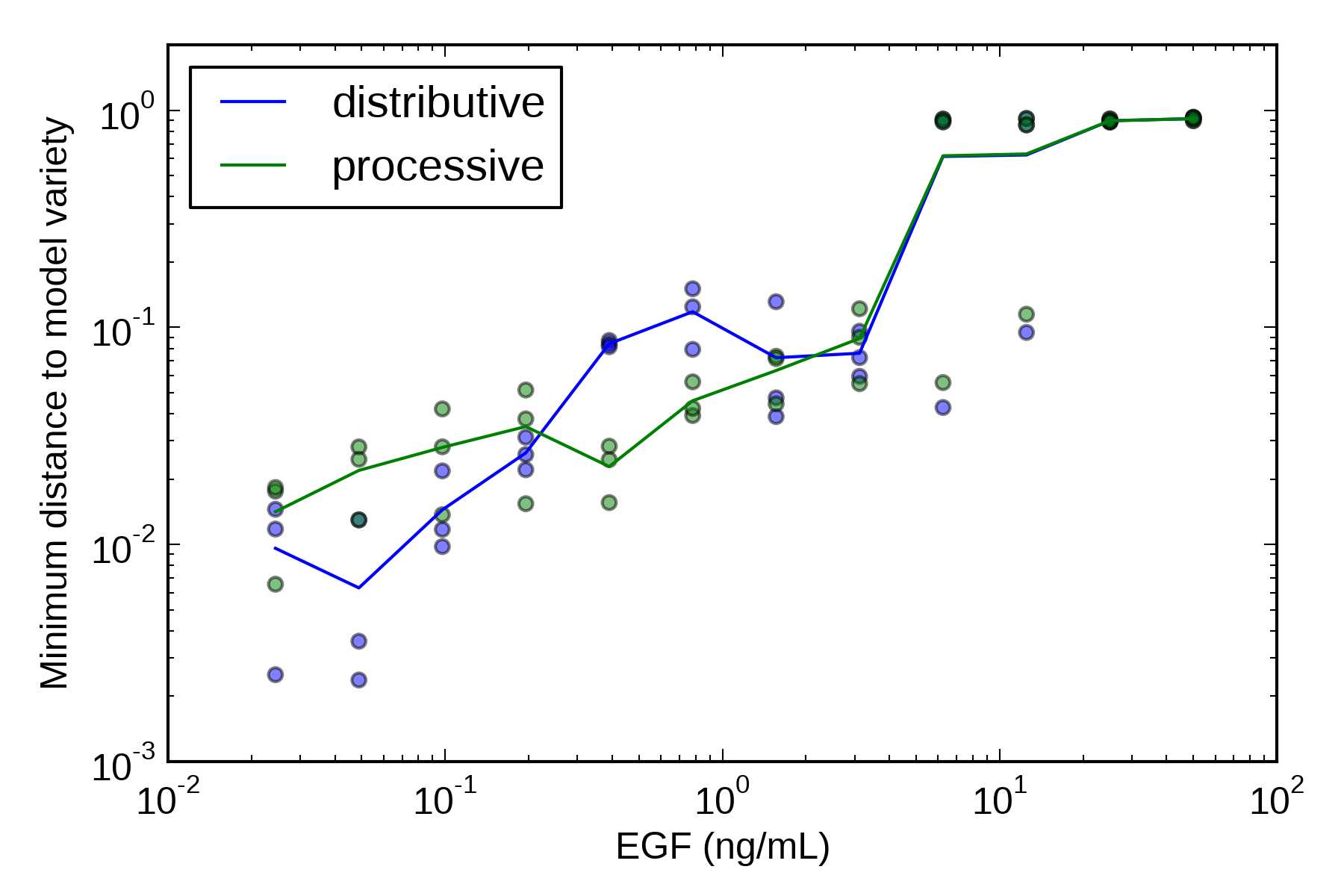}
\caption{MAP Kinase model selection using 36 data points from 3 model variables (aggregate phosphoforms of ERK) taken in triplicate at each of 12 EGF stimulation levels.}
\label{fig-lda}
\end{figure}

\bibliographystyle{pnas}
\bibliography{references}

\end{document}




\title{Supporting Information for: \\Numerical algebraic geometry for model selection and its application to the life sciences}

\author{Elizabeth Gross$^{1}$, Brent R. Davis$^{2}$, Kenneth L. Ho$^{3}$, Daniel J. Bates$^{2}$, Heather A. Harrington$^{4}$ \\
$^{1}${Department of Mathematics, San Jos\'{e} State University}, 
\\
$^{2}${Department of Mathematics, Colorado State University},
\\
$^{3}${Department of Mathematics, Stanford University},
\\
$^{4}$Mathematical Institute, University of Oxford.}
\date{} 
\maketitle
\tableofcontents


\section{Geometry}
In this section we briefly review the some of the fundamental geometric concepts needed for the methods in the main text. \\
\\
\subsection{Numerical algebraic geometry for isolated solutions}
{\em Numerical algebraic geometry} refers to the use of numerical methods, particularly homotopy continuation-based methods, to compute approximations to 
solutions of polynomial systems.  In other words, given a polynomial system $\bS{F}:\mathbb C^N\to\mathbb C^n$ with $n$ equations in $N$ variables, numerical 
algebraic geometry seeks to find numerical approximations of all $\bS{z} \in \C^N$ such that $\bS{F}(\bS{z})=\bS{0}$.  It may be the case that the solution set of $\bS{F}$ has infinitely many such points (curves, surfaces, 
etc.), in which case the data structure that encodes the solutions is called a \emph{witness set}.  See the description of the {\em numerical irreducible decomposition} below for more on this.   For simplicity, we assume in this section that $N=n$. 
The books~\cite{SW,SIAMbook} and the references therein provide much more detailed explanations than those included in this section, from more 
mathematical and computational perspectives.  

The core technique for most numerical algebraic geometry algorithms is {\em homotopy continuation}.  The idea of homotopy continuation is to cast the polynomial system 
to be solved, say  $\bS{F}:\mathbb C^N\to\mathbb C^N$, as a member of a parameterized family of polynomial systems, $\bS{H}:\mathbb C^N\times\mathbb C\to\mathbb C^N$,  called 
a homotopy $\bS{H}(\bS{z},t)$ with parameter $t\in\mathbb C$, that includes one polynomial system $\bS{G}:\mathbb C^N\to\mathbb C^N$ that is easily solved and has the special property of having ``enough'' isolated solutions.  In this document, we use the {\tt Bertini} standard assumption that $\bS{H}(\bS{z},1)=\bS{G}(\bS{z})$ and $\bS{H}(\bS{z},0)=\bS{F}(\bS{z})$, i.e., $t$ marches from 1 to 0.  There are several canonical options for the construction of such a homotopy, and the reader is encouraged to consult the general references above and the references therein for further details.

As $t$ varies from 1 to 0, some results from algebraic geometry tell us that the solutions of the polynomial system $\bS{H}(\bS{z},t)=\bS{0}$ vary continuously and generally stay distinct until $t=0$, where they may converge to solutions of $\bS{F}(\bS{z})=\bS{0}$ or diverge.  More specifically, there is a measure zero subset of $t\in\mathbb C$, meaning a finite set of points in this particular parameter space, over which two or more solutions coalesce.  Such occurrences are thus probability zero events and, furthermore, can be detected on the fly and avoided.  Said 
more technically, there is a {\em Zariski open, dense} set of the parameter space above which the solution set is finite and consists of a fixed number of solutions.  Here, ``Zariski'' refers to the {\em Zariski topology}, for which basic open sets are the complements of solution sets of polynomial systems.

In practice, $t$ is moved in discrete increments, not continuously.  For each solution at $t=1$, a path of solutions is {\em tracked} using numerical predictor/corrector methods as $t$ advances to 0.  Implementations typically utilize adaptive step lengths and adapative precision.  There are far too many details about this procedure to give a thorough explanation here.  Instead, refer to the references above (and those therein) for further details.

Ultimately, the output of this procedure is a superset of numerical approximations of the isolated solutions of $\bS{F}(\bS{z})=\bS{0}$, possibly including approximations to points lying on positive-dimensional components, if any.  It is important to note that this procedure necessarily works over $\mathbb C$ and finds all complex solutions.  Real solutions could be 
buried somewhere within the complex solutions, and it is particularly difficult to extract these outside of the zero-dimensional case.  However, methods do exist to extract such a real point \cite{Wu13,Rouillier,hauenstReal}; here we use the method of~\cite{hauenstReal}, which is guaranteed (with probability one) to minimize the distance of a prescribed real point to each real connected component of the variety defined by $\bS{F}(\bS{z}) = \bS{0}$.

\subsection{Numerical algebraic geometry for positive-dimensional solution sets}

For solution sets of positive dimension (curves, surfaces, etc.), there is an extension of homotopy continuation referred to as the {\em numerical irreducible decomposition (NID)}.  
As opposed to the case of systems of linear equations (at most one solution component of one dimension), there may be many components of many different dimensions.  For example, one solution set might consist of seven components of dimension four, five surfaces, three curves and 15 isolated points.  Furthermore, components may be singular, meaning that the Jacobian matrix is rank-deficient throughout the component.

Technical definitions of dimension, {\em irreducible component}, and the like go a bit beyond the scope of this paper.  It is enough to know that each ``piece'' of a solution set has a fixed dimension (e.g., a curve has dimension one, a surface two, etc.) and by the {\em dimension} of a solution set of a polynomial system of equations, we mean the maximum of the dimensions of the irreducible components.

The {\em numerical irreducible decomposition} of a solution set of a polynomial system consists of a catalog of the dimensions and degrees of each irreducible component, along 
with a set of {\em witness points} on each component.  By {\em degree}, we mean the number of points in the intersection of a component with a randomly-chosen affine linear 
space of complementary dimension.  The {\em witness points} on a component are then exactly these points (and thus depend on the choice of linear space).  One fundamental 
result from algebraic geometry is that an irreducible component will {\em almost} always intersect a complementary-dimensional linear space exactly in a set of points {\em and} that 
the number of points is the same for {\em almost} any choice of linear space.  Again, this can be stated as a probability one guarantee or with Zariski open, dense sets, but we choose not to be that technical here.

\subsection{Software for numerical algebraic geometry}

Various numerical algebraic geometry software packages have been produced over the years.  Currently, there are three main options:  PHCpack~\cite{PHCpack}, 
HOM4PS-3~\cite{HOM4PS}, and Bertini~\cite{bertini}, each with their own benefits and drawbacks.  In this article, we used Bertini exclusively.
\\

\subsection{Geometry specific to presented algorithms}\label{subsec:geometry}

Theorem 1 of the main text should be familiar to those trained in multivariate calculus; this is essentially the method of Lagrange multipliers.  Geometrically, this system of 
equations forces the normal directions of the objective function and the constraints to line up in the same direction (up to some scalar(s), the Lagrange multiplier(s)). 

When working with an irreducible component of the solution set of a system of polynomial equations, it is often useful to deal with a {\em complete intersection}.  Said 
simply, the idea is that computations can be more difficult if there are more equations than necessary.  To see why this might be true, let us consider an example from 
linear algebra.  Suppose we have a single linear equation in three variables, defining a plane.  Now suppose we consider a system of two equations consisting of 
that equation and twice that equation (having the same solution set).  Then the matrix of coefficients of this linear system is not full rank (not a desirable situation) and 
we have two equations defining a geometric object that could be described by a single equation.

In the nonlinear setting, the situation is quite similar.  Having ``too many'' equations leads to an undesirable rank-deficient Jacobian matrix.  Suppose polynomial system 
$\bS{F}:\C^N \to \C^n$ has an irreducible component $X$ of dimension $N-m$ ({\em codimension} $m$).  Then, again with probability one, the polynomial system 
$\bS{A}\cdot \bS{F}$ has $X$ as an irreducible component but has the ``correct'' number of equations, where $A$ is a random constant matrix with $n$ columns and $m$ rows.  Here, ``correct'' means the number of equations matches to codimension $m$.  We will refer to this method as \emph{squaring up}.

Finally for this section specific to the algorithms developed in the main text, we require the user to check two geometric facts, the meaning of which may not be entirely clear. 
\begin{enumerate}
\item $\mathcal V_{\cM} \cap \mathcal V_{\cD}$ 
refers to the intersection of the model and data varieties, as defined in the main text.  To find the intersection of two solution sets, it is sufficient to simply solve the system consisting  
of all equations appearing in the systems for $\mathcal V_{\cM}$ and $\mathcal V_{\cD}$, i.e., the union of those two polynomial systems.  There are more sophisticated methods, but this is 
sufficient.  

\item The user must determine whether the intersection of $(\mathcal V_{M})_{\R}$ and $(\mathcal V_{M})_{\R}$ is empty.  By this, we simply mean that one should search for real points 
in the intersection just described, e.g., using the method of~\cite{hauenstReal}.
\end{enumerate}

\section{Choice of test statistics and parameter estimates}
In this section, we justify our procedures for model validation and parameter estimation.

\subsection{Maximum likelihood}

Here we justify the assertion that the test statistic given in the main text is related to likelihood maximization. 

Consider first, for simplicity, the case of a single data point $\hat y=(\hat y_1, \ldots, \hat y_m)$, which we assume is a perturbation $\hat y= \xi + \epsilon$ of some unknown true value $\xi = (\xi_{1}, \dots, \xi_{m})$, where each component $\epsilon_{i}$ of the error $\epsilon = (\epsilon_{1}, \dots, \epsilon_{m})$ is an independent zero-mean Gaussian random variable with variance $\sigma_{i}^{2}$. We are interested in computing the probability that $\hat y$ comes from a given model as defined by a model variety $\mathcal V_{\cM}$. A point on $\mathcal V_{\cM}$ has the form $(a, x, y)$, where $a=(a_1, \ldots, a_k)$ are the model parameter values, $x=(x_1, \ldots, x_n)$ are the variable values, and $y=(y_1, \ldots, y_n)$ are the outputs.
  The probability that $\hat y$ comes from a given point $(a, x, y) \in \mathcal{V}_{\cM}$, i.e., that $\hat y$ is a perturbation of $y$ where $(a, x, y) \in \mathcal{V}_{\cM}$ for some $a$ and $x$, is then
\begin{align*}
 \Pr (\hat y\: | \: a, x, y) = \Pr (\hat y\: | \: \xi= y) = \prod_{i = 1}^{m} \Pr (\hat{y}_{i} \: | \: \xi_{i} = y_{i}).
\end{align*}
This is also called the likelihood $L(a, x, y)$ of $(a, x, y)$ and we wish to find its maximizer over all $(a, x, y) \in \mathcal{V}_{\cM}$. This can equivalently be done by considering the log-likelihood, which gives
\begin{align*}
 \log L(a, x, y) = \sum_{i = 1}^{m} \log \Pr (\hat{y}_{i} \: | \: \xi_{i} = y_{i}) = \sum_{i = 1}^{m} \left( \frac{1}{2} \log 2 \pi \sigma_{i}^{2} - \frac{(\hat{y}_{i} - y_{i})^{2}}{2 \sigma_{i}^{2}} \right)
\end{align*}
by normality. The maximizer $(a^{*}, x^{*}, y^{*})$ can therefore be found by solving the optimization problem
\begin{align*}
 d^2 = \min_{(a, x, y) \in \mathcal V_{\cM}} \sum_{i = 1}^{m} \frac{(\hat{y}_{i} - y_{i})^{2}}{\sigma_{i}^{2}},
\end{align*}
where the optimum is precisely the test statistic. The values $a^{*}$, $x^{*}$, and $y^{*}$ are the maximum likelihood estimates for, respectively, the parameters (estimation), the unobservable variable values (inference/recovery), and the true output values (filtering/denoising), 

The test statistic $d^2$ itself also has a useful interpretation as follows. Suppose that $\hat y $ comes from a point $(a, x, y) \in \mathcal V_{\cM}$. Then
\begin{align*}
 d^2 = \sum_{i = 1}^{m} \frac{(\hat{y}_{i} - y^{*}_{i})^{2}}{\sigma_{i}^{2}} \leq \sum_{i = 1}^{m} \frac{(\hat{y}_{i} - y_{i})^{2}}{\sigma_{i}^{2}}
\end{align*}
by definition. But regarding each $\hat{y}_{i}$ as a random variable, each term $(\hat{y}_{i} - y_{i}) / \sigma_{i}$ in the summation above is standard normal. Hence the right-hand side has a chi-squared distribution with $m$ degrees of freedom ($\chi^{2}_{m}$). The inequality should be interpreted by regarding $d^2$ as a random variable subject to the same source of randomness. This can be written somewhat clearer as
\begin{align*}
 d^2(\omega) \leq \sum_{i = 1}^{m} \frac{(\hat{y}_{i} (\omega) - y_{i})^{2}}{\sigma_{i}^{2}},
\end{align*}
where we have made explicit the underlying dependence of both sides on the same random realization $\omega$. The inequality then holds for each value of $\omega$. Consequently, we conclude that
\begin{align*}
 \Pr (d^2 \leq u) \geq \Pr (U \leq u), \quad U \sim \chi^{2}_{m},
\end{align*}
so
\begin{align*}
 \Pr (d^2 \geq p_{\alpha}) \leq \Pr (U \geq p_{\alpha}) = \alpha, \quad U \sim \chi^{2}_{m},
\end{align*}
where $p_{\alpha}$ is the upper $\alpha$-percentile for $\chi^{2}_{m}$. This can be used to test the hypothesis that $\hat y$ comes from $\mathcal V_{\cM}$. The test statistic is also related to the log--maximum-likelihood as
\begin{align*}
 \log L(a^{*}, x^{*}, y^{*}) = \frac{1}{2} \left( m \log 2 \pi + \sum_{i = 1}^{m} \log \sigma_{i}^{2} - d^2 \right),
\end{align*}
which is a useful quantity for model selection via, e.g., the Akaike or Bayesian information criteria.

Now consider the case of multiple data points $\{ \hat y^{(j)} \}_{j = 1}^{p}$. As before, we assume that each $\hat y^{(j)} = (\hat y^{(j)}_1, \dots, \hat y^{(j)}_m)$ is a perturbation $\hat y^{(j)} =\xi^{(j)}+ \epsilon^{(j)}$, where each $\epsilon^{(j)}_{i}$ is an independent zero-mean Gaussian random variable with variance $\sigma_{j,i}^{2}$. Instead of searching for one point on $\mathcal V_{\cM}$, we now have to search for $p$ points $(a, x^{(j)}, y^{(j)})$ for $j = 1, \dots, p$ all with the same parameter values (since they come from the same fixed model realization). The probability that $\hat y^{(j)}$ comes from $(a, x^{(j)}, y^{(j)})$ for $j = 1, \dots, p$ is then
\begin{multline*}
 \Pr (\hat y^{(1)}, \dots, \hat y^{(p)}\: \vert \: a, x^{(1)}, y^{(1)}, \dots, x^{(p)},y^{(p)}) = \prod_{j = 1}^{p} \Pr (\hat y^{(j)} \: \vert \: \xi^{(j)} = y^{(j)}) = \prod_{j = 1}^{p} \prod_{i = 1}^{m} \Pr (\hat{x}_{j,i} \: \vert \: \xi^{(j)}_{i} = y^{(j)}_{i})\\
 \equiv L(a, x^{(1)}, y^{(1)}, \dots, x^{(p)},y^{(p)})
\end{multline*}
by independence. This is essentially the same as before except that we now loop over each coordinate of each data point. Therefore, the maximum likelihood estimates $(a^{*}, x^{(1)^*}, y^{(1)^*}, \dots, x^{(p)^*},y^{(p)^*})$ can be obtained by solving
\begin{align*}
 d^2 = \min_{(a, x^{(1)}, y^{(1)}, \dots, x^{(p)},y^{(p)}) \in \mathcal V_{\cM}} \sum_{j = 1}^{p} \sum_{i = 1}^{m} \frac{(\hat{y}^{(j)}_{i} - y^{(j)}_{i})^{2}}{\sigma_{j,i}^{2}}.
\end{align*}
The same arguments go through and we find that $\Pr (d^2 \geq p_{\alpha}) \leq \alpha$ for $p_{\alpha}$ the upper $\alpha$-percentile for $\chi^{2}_{mp}$. The log--maximum-likelihood is related to $d^2$ as
\begin{align*}
 \log L(a^{*}, x^{(1)^*}, y^{(1)^*}, \dots, x^{(p)^*},y^{(p)^*}) = \frac{1}{2} \left( pm \log 2 \pi + \sum_{j = 1}^{p} \sum_{i = 1}^{m} \log \sigma_{j,i}^{2} - y \right).
\end{align*}

\section{Algorithm modifications}

The algorithms  presented in the main text are in their simplest form.  Some applications  require modifications, particularly if there are constraints on the variables or parameters.

\subsection{Solving the constrained optimization problem} \label{constraints} \label{subset:boundary}

In many common settings, there exist constraints on the variable and parameter spaces.  For example, in chemical reaction networks, the rate parameters are all assumed to be positive. Thus,  when positivity or other constraints are present, instead of finding the weighted squared distance between two varieties, we are finding the weighted squared distance between two \emph{semi-algebraic sets}, i.e. sets defined by polynomial equalities and inequalities as opposed to just polynomial equalities. Indeed, if we let $S_{\mathcal M} \subset (\mathcal {V_M})_{\R}$ denote the semi-algebraic set associated to the model, e.g. $S_{\mathcal M}=\mathcal {V_M} \cap \R_{\geq 0}^{k+n+m}$, then the appropriate statistic is:

 $$d^2=\min  \sum_{i=1}^m \frac{(y_i - \hat y_i)^2}{\sigma_i^2} \label{eq:modelCompMin}
 \text{ subject to }  (a, x, y) \in S_{\mathcal M}.$$

In the case when a bound on the statistic $d^2$ is sufficient, then no additional work is needed.  One can solve the system from Proposition 1, keeping in mind that the weighted squared distance between the closest pairs of points returned would be a lower bound on $d^2$.  If the closest point to $(\mathcal {V_D})_{\R}$ in $(\mathcal {V_M})_{\R}$ is also an element of $S_{\mathcal M}$, then the squared distance would be exactly the statistic $d^2$.

When the exact value of $d^2$ is needed, then one should solve the Karush-Kuhn-Tucker (KKT)  system of equations. Let $F_1, \ldots, F_r$, $h_1, \ldots, h_s$ be polynomials in the ring $\R[a_1, \ldots, a_k,  x_1,\ldots, x_n, y_1, \ldots, y_m]$.  Let $S_{\mathcal M}$ be the semi-algebraic set of all $(a, x, y) \in \R^{k+n+m}$ that satisfies
\begin{align*}
F_i(a, x, y) =0 & \text{ for } i=1, \ldots, r \\
h_i(a, x, y) \leq 0 & \text{ for } i=1, \ldots, s
\end{align*}
Let $ \lambda_1, \ldots, \lambda_r, \mu_1, \ldots, \mu_s$ be indeterminates (these are the KKT multipliers). The KKT system is
\begin{align}
f& =0 \label{eq:KKT1}\\
\lambda_1 \nabla f_1 + \ldots \lambda_r \nabla f_r + \mu_1 \nabla h_1 + \ldots \mu_s \nabla h_s + \left(\begin{array}{c}0 \\y - \hat y\end{array}\right) & =0 \label{eq:KKT2}\\
\mu_1 h_1 & = 0 \\
&\vdots \\
\mu_s h_s & =0 . \label{eq:KKT5}
\end{align}

In order for $(a, x, y)$ to be a critical point, the point $(a, x, y)$ must be a solution to this system and satisfy the inequalities $h_i \leq 0$ for $i=1, \ldots, s$ and $\mu_i \geq 0$ for $i=1, \ldots, s$.  Thus, we can find the global minimum by using numerical algebraic geometry to solve the system defined by equations \eqref{eq:KKT1} - \eqref{eq:KKT5}, then filtering the solutions appropriately.  This combination of numerical algebraic geometry and the KKT equations was first employed in~\cite{RFBBM}.

Alternatively, in some situations, it may be more efficient to minimize the objective function over $(\mathcal{V_M})_{\R}$ and then check the boundary of $\mathcal S_{\mathcal M}$. We describe this method for when there are non-negative constraints on all indeterminates. After solving the constrained optimization problem on $(\mathcal{V_M})_{\R}$, we assume, in an orderly fashion, that one of the indeterminates, say $x_{i}$, is zero. In this case, we are required to minimize the distance of $\mathcal V_{\cM} \cap \{ x_i = 0\}$ to $\mathcal V_{\cD}$. This involves solving a smaller and related constrained optimization problem. In total, if there are $N$ indeterminates, there are $2^N - 1$ combinations to set to zero. This amount to solving $2^N - 1$ Lagrange multiplier systems.  

One observation is that the number of systems that need to be solved explodes when $N$ is large. Note however that the dimension of $\mathcal V_{\cM} \cap \mathcal \{x_{i}=0\}$ is less than or equal to the dimension of $\mathcal V_{\cM}$, with the inequality being strict when $V_{\cM}\subsetneq \{ x_i = 0\}$. Geometrically, there are no longer degrees of freedom in the variable $x_{i}$ so the dimension is reduced. Furthermore, we can expect the dimension to be reduced by one.  As we impose additional constrains, $x_{j} = 0$ for $i \neq j$, the dimension may drop further.

In practice, there are diminishing returns as you begin setting $x_{i}$ to zero.  That is, there exists some positive integer $k$ such that $\mathcal V_{\cM} \cap \{x_{i_1}=\ldots=x_{i_k}=0\}$ is zero dimensional for some indexing set $i_1 < \cdots < i_{k}$. In this case, measuring the distance of $\mathcal V_{\cM} \cap \{x_{i_1}=\ldots=x_{i_k}=0\}$ to $\mathcal V_{\cD}$ using a  Lagrange multiplier method is unnecessary, and,  
as we continue imposing additional constraints on $\mathcal V_{\cM}$, either the intersection is non-empty and every observable variable is set to zero, or the model variety becomes empty. Thus, it becomes redundant or unnecessary to check additional cases.

For example, in Section \ref{subsect:MAPKopt}, the MAPK model variety $\mathcal V_{\cM}$ is one-dimensional and $\mathcal V_{\cM} \cap \{x_i=0\}$ becomes zero-dimensional for each $i$.  For the cases where $x_i = x_j = 0$ for $i \neq j$, the intersection of the corresponding linear spaces with $\mathcal V_{\cM}$ becomes empty. Even though there are $2^{16} - 1 = 65,535$ boundary cases to check, there are really only 16 relevant cases.  See Section \ref{subsect:MAPKopt} for more explicit details on how this calculation carried out.

\subsection{Removing extinction components}\label{sec:extinction}

Given a model, it is quite common that the model variety is not irreducible but instead is the union of several irreducible components.  In applications, it may be preferred to remove from consideration components that lie entirely in a coordinate hyperplane, since, in such components, one or several of the parameters and/or variables are equal to zero.  For example, in a chemical reaction network, such a component is called an \emph{extinction component} \cite{GHRS15} since it captures the situation where one or more of the reactants have ``run out." It is common to want to avoid extinction components when estimating parameters.

Removing components where a parameter or variable is equal to zero throughout the set from consideration can be done algebraically with \emph{saturation}.  In particular, if $I_{\mathcal M}$ is the defining ideal of the model $\mathcal V_{\mathcal M}$ one should compute
$$ I_{Main}=I_{\mathcal{M}} \ : \ (a_1\cdots a_k\cdot x_1 \cdots x_n \cdot y_1 \cdots y_m)^{\infty}:=$$
$$\{ f \in \R[a_1, \ldots, a_k,  x_1,\ldots, x_n, y_1, \ldots, y_m] \ : \  \exists k \in \mathbb{N} \text{ s.t. } (a_1\cdots a_k\cdot x_1 \cdots x_n\cdot y_1 \cdots y_m)^k f \in I_{\mathcal{M}} \}.$$
This procedure can be performed using the {\tt saturate} command in {\tt Macaulay2}.

To estimate parameters such that the best estimate corresponds to a point not on an extinction component of $\mathcal V_{\mathcal M}$, one should modify Algorithm 3, replacing $\mathcal V_{\mathcal M}$ with $\mathcal V(I_{Main})$.

\section{Results}
We provide details for the calculations of examples in the main text. All code is available at:\\ {\tt http://www.math.sjsu.edu/~egross/NAGModelSelection/AuxillaryFiles}.

\subsection{Cell death activation}\label{subsec:celldeath}

We provide the details of the calculations for model compatibility of the cell death cluster model.  This subsection includes detailed information regarding the solving schemes available in NAG software that were utilized.  A summary for the practioner can be found at the end of the subsection.

The model describes activation of apoptosis by death receptor Fas mechanisms \cite{scott:nat:2009}.  The model includes constitutive receptor opening and closing, pairwise open Fas stabilization, higher-order open Fas stabilization enabled by FasL, and ligand-induced receptor opening. According to its conformational states, Fas is assumed to be one of three species: closed ($X_1$); open, unstable ($X_2$); and open, stable ($X_3$), i.e., active and signaling. Furthermore, let the ligand FasL be denoted by $L$. Then the model has the reactions

 \begin{align*}
  X_2 &\cee{<=>[k_{c}][k_{o}]} X_1,\\
  X_3 &\cee{->[k_{u}]} X_2,\\
  jX_2 + (i - j)X_3 &\cee{->[k_{s}^{(i)}]} (j - k)X_2 + (i - j + k)X_3,\\
  L + jX_2 + (i - j)X_3 &\cee{->[k_{l}^{(i)}]} (j - k)X_2 + (i - j + k)X_3
 \end{align*}
for $i \in \{ 2, 3 \}$, $j = 1, \dots, i$, and $k = 1, \dots, j$. The first reaction describes spontaneous receptor opening and closing. The second reaction describes constitutive destabilization of open Fas. The third reaction describes cluster-stabilization by open Fas, independent of the presence of FasL. The fourth reaction describes cluster-stabilization events enabled by FasL.

Assuming mass-action kinetics, the reactions can be translated as follows:
\begin{align*}
\begin{cases}
\dot{x_1} &= -v_1, \\
\dot{x_2} &= v_1+ v_2 - v_3 - v_4, \\
\dot{x_3} &= v_3+v_4-v_2,
\end{cases}
\quad \text{where \;}
\begin{cases} v_1&=k_ox_1 + (-k_c)x_2, \\
 v_2&=k_ux_3, \\
v_3&= 6 k_s^{(3)} x_2^3 + 3 k_s^{(3)} x_2^2 x_3 + 3 k_s^{(2)} x_2^2 +
k_s^{(3)} x_2 x_3^2 + k_s^{(2)} x_2x_3,\\
v_4&= 6 k_l^{(3)} x_2^3 l + 3 k_l^{(3)} x_2^2 x_3l + 3 k_l^{(2)}
x_2^2 l + k_l^{(3)} x_2 x_3^2 l + k_l^{(2)} x_2x_3l,
\end{cases}
\end{align*}
where $v_i$ are the reaction velocities for the variables $x_i$, and lowercase letters denote the concentrations of their uppercase counterparts.

The model parameters for the cell death cluster model are
$${a} = (k_o,k_c,k_u,k_s^{(2)},k_s^{(3)},k_l^{(2)},k_l^{(3)})$$ 
and the variables are
$${x} = (\ell,x_1,x_2,x_3)$$
The outputs are
$$y=(\lambda,\rho,\zeta)$$
representing, respectively, the total ligand concentration, the total receptor concentration, and the total downstream ``death signal'', as given by the equations
\begin{eqnarray}
\lambda - \ell &=& 0 \label{eqn:clusterdata1} \\
\rho - (x_1 + x_2 + x_3) &=& 0 \\
\zeta - x_3 &=& 0. \label{eqn:clusterdata3} \
\end{eqnarray}
We set the model equations ($\dot{x}_1,\dot{x}_2,\dot{x}_3$) to zero, and, together with equations \eqref{eqn:clusterdata1}--\eqref{eqn:clusterdata3}, we obtain the defining equations for the model variety $\mathcal V_{\cM}$.  The ambient space that $\mathcal V_{\cM}$ is contained in has dimension $14$. This space has coordinates defined by both the model parameters $\bS{a}$, the variables $\bS{x}$ and the outputs $y$.  


Given an observable data point $\hat y = (\widehat{\lambda}, \widehat{\rho}, \widehat{\zeta})$, we define the data variety as: 
\begin{eqnarray}
\mathcal V_{\cD} &=& \{(x,a,y) \in \C^{14} : \lambda = \widehat{\lambda}, \, \rho = \widehat{\rho}, \, \zeta = \widehat{\zeta} \} \label{set:datavariety1} \
\end{eqnarray}
for the clustering model.  Note that $\mathcal V_{\cD}$ has dimension $11$ since there are no degrees of freedom in the variables $\lambda, \rho$, or $\zeta$.

We first compute a numerical irreducible decomposition (NID) of $\mathcal V_{\cM}$ using {\tt Bertini}; this will aid in understanding $\mathcal{V_M} \cap \mathcal{V_D}$. One can verify, after computing the NID for $\mathcal V_{\cM}$, that $\mathcal V_{\cM}$ is a $9$-dimensional complex algebraic set of degree $10$ (file name:  {\tt Cluster\_Model\_NID} ). 

Now suppose we are given the following data point (taken from the model without noise): 
$$\hat y = (\widehat{\lambda}, \widehat{\rho}, \widehat{\zeta}) = (1.7784308, 2.31883024, 2.16896112).$$ 
One can then verify using the NID that $\mathcal V_{\cM} \cap \mathcal V_{\cD} \neq \emptyset$ (file name: {\tt Cluster\_Model\_Data\_NID}).  That is, the intersection of the model variety and data variety is nonempty. Specifically, $\mathcal V_{\cM} \cap \mathcal V_{\cD}$ is a $6$-dimensional complex algebraic set of degree $5$. Adding noise to the coordinates of $\hat y$ taken from $\mathcal N(0,0.1)$ did not affect the dimension or degree. Again, we got that $\mathcal V_{\cM} \cap \mathcal V_{\cD}$ has dimension 6 and degree 5.

Since we are interested in model compatibility, our goal is to find at least one nonnegative point in $(\mathcal V_{\cM})_{\R} \cap (\mathcal V_{\cD})_{\R}$.  The above computation at least provides evidence that this is the case, but it may be possible that $\mathcal V_{\cM} \cap \mathcal V_{\cD}$ does not contain any nonnegative real points or any real points at all for that matter.  We will approach this problem using the methods described in \cite{hauenstReal}. If $\mathcal V_{\cM} \cap \mathcal V_{\cD}$ contains a real nonnegative point then we cannot reject the model and may conclude that the model is compatible with the data.  If $(\mathcal V_{\cM})_{\R} \cap (\mathcal V_{\cD})_{\R}$ does not contain a real point then we can try and use a more general Lagrange multiplier method similar to the one employed in Section \ref{subsect:MAPKopt} in dealing with model selection.

We first randomly select a real, positive point:
\begin{eqnarray*}
\ell &=& 6.491154749564521 \\
x_1 &=& 7.317223856586703 \\
x_2 &=& 6.477459631363067 \\
x_3 &=& 4.509237064309449 \\
k_{o} &=& 5.470088922863450 \\
k_{c} &=& 2.963208056077732 \\
k_{u} &=& 7.446928070741562  \\
k_s^{(2)} &=& 1.889550150325445 \\
k_s^{(3)} &=& 6.867754333653150 \\
k_l^{(2)} &=& 1.835111557372697 \\
k_l^{(3)} &=& 3.684845964903365
\end{eqnarray*}
where each coordinate is chosen uniformly on the interval $[0,10]$. This point will determine the observable, i.e. output, variables $\lambda,\rho$, and $\zeta$ using equations \eqref{eqn:clusterdata1}--\eqref{eqn:clusterdata3}. Call this point $(a^{\star}, x^{\star}, y^{\star}) \in \R^{14} $. For the time being, the left and right endpoints of each subinterval $[0,10]$ have been chosen arbitrarily. It is unclear, at this time, how the endpoints or length of the interval affects the performance in finding nonnegative real points contained in $(\mathcal V_{\cM})_{\R} \cap (\mathcal V_{\cD})_{\R}$ using the methods described below.  

Our aim then is to solve the constrained optimization problem:
\begin{equation}
\label{opt:constrainedhaun}
\begin{aligned}
& \underset{\bS{y}}{\text{minimize}}
& & \| (a, x, y) - (a^{\star}, x^{\star}, y^{\star}) \|^2 \\
& \text{subject to}
& & \bS{(a, x, y)} \in (\mathcal V_{\cM})_{\R} \cap (\mathcal V_{\cD})_{\R}.
\end{aligned}
\end{equation}
Geometrically, we are minimizing the distance between the chosen point $(a^{\star}, x^{\star}, y^{\star})$ and $(\mathcal V_{\cM})_{\R} \cap (\mathcal V_{\cD})_{\R}$.  

We will refer to the system defining $(\mathcal V_{\cM}) \cap (\mathcal V_{\cD})$ as $\bS{f^*}(\bS{a,x,y})$. Squaring up the polynomial system will be a necessary step when utilizing the perturbed regenerative solving scheme. 
Squaring up was briefly described in Section \ref{subsec:geometry} but we will give additional details here.  First notice from previous computations that $\mathcal V_{\cM} \cap \mathcal V_{\cD}$ has codimension $14-6 = 8$.  Thus, there exists a nonempty Zariski open set $\mathcal{A} \subseteq \C^{8 \times 9}$ such that for every matrix $\bS{A} \in \mathcal{A}$, we have $\V_{\cM} \cap \V_{\cD} \subseteq \V(\bS{Af^*}(\bS{a,x,y}))$. This means we may take $8$ random $\C$-linear combinations of the equations defining $\mathcal V_{\cM} \cap \mathcal V_{\cD}$ and, with probability one, still cut out at least $\V_{\cM} \cap \V_{\cD}$.  It is sufficient to sample the elements of $\bS{A}$ uniformly along the complex unit circle.  

As a side note, we may take additional steps to reduce complexity by replacing the matrices $\bS{A}$ with matrices of the form $[\bS{I}_{8} \,\, | \,\, \bS{b}] \subseteq \mathcal{A}$ where $\bS{I}_{8}$ denotes the $8 \times 8$ identity matrix and $\bS{b}$ denotes a $8 \times 1$ column vector who elements are sampled uniformly along the complex unit circle.  This has the effect of adding a random multiple of the last function to each of the other functions. One may verify \emph{a posteriori} if a point $\bS{x} \in \mathcal V(\bS{Af^*}(\bS{x}))$ is also in $\mathcal V_{\cM} \cap \mathcal V_{\cD}$ by function evaluation of $\bS{f^*}(\bS{a,x,y})$. 

The polynomial system for the optimization problem \eqref{opt:constrainedhaun} is:
\begin{eqnarray}
h_{i}(a,x,y) &=& 0, \mbox{ for } 1 \leq i \leq 8, \mbox{ where } h_{j}(a,x,y) = \Sigma_{i=1}^9 A_{jk} f^*_{k}(a,x,y) \mbox{ with } A_{jk} = [\bS{A}]_{j,k} \label{eq:lagrangecluster1} \\
a_i - a^{\star}_i &=& \Sigma_{j=1}^{8} \frac{\partial h_{j}(a,x,y)}{\partial a_{i}} \lambda_{j}, \mbox{ for } 1 \leq i \leq 7 \label{eq:lagrangecluster2} \\
x_i - x^{\star}_i &=& \Sigma_{j=1}^{8} \frac{\partial h_{j}(a,x,y)}{\partial x_{i}} \lambda_{j}, \mbox{ for } 1 \leq i \leq 4 \label{eq:lagrangecluster3} \\
y_i - y^{\star}_i &=& \Sigma_{j=1}^{8} \frac{\partial h_{j}(a,x,y)}{\partial y_{i}} \lambda_{j}, \mbox{ for } 1 \leq i \leq 3. \label{eq:lagrangecluster4} \
\end{eqnarray}
We call $\bS{\lambda} = (\lambda_1,\ldots,\lambda_{8})$ the Lagrange multipliers for $\mathcal V_{\cM} \cap \mathcal V_{\cD}$.  This is a system of $22$ variables and $22$ equations.  Written more compactly we may write equations \eqref{eq:lagrangecluster1}--\eqref{eq:lagrangecluster4} as:
\begin{eqnarray}
\bS{h}(\bS{a,x,y}) &=& \bS{0} \label{eq:lagrangeclustersimple1} \\
\bS{(a,x,y)}^{T} - {(a^{\star}, \bS{x}^{\star}, y^{\star})}^{T} &=& \bS{J}^{T}_{\bS{h}}(\bS{a,x,y})\bS{\lambda}^{T} \label{eq:lagrangeclustersimple2} \
\end{eqnarray}
where $\bS{J}_{\bS{h}}(\bS{a,x,y})$ is the Jacobian matrix of $\bS{h}(\bS{a,x,y})$. The Lagrange multipliers in equations \eqref{eq:lagrangeclustersimple2} need not be real since we are taking $\C$-linear combinations of $\bS{f}^*(\bS{a,x,y})$. 

Regeneration is a numerical algebraic geometry method we found to be most relevant to solve equations \eqref{eq:lagrangeclustersimple1}--\eqref{eq:lagrangeclustersimple2} and is implemented in {\tt Bertini}.  Regeneration uses a linear product homotopy scheme in which each equation is built up one at a time.  Depending on the ordering and structure of each equation, that can lead to huge computational savings.  However, regeneration is restrictive in that it may not find all \emph{singular} isolated solutions to equations \eqref{eq:lagrangeclustersimple1}--\eqref{eq:lagrangeclustersimple2}.  For the finer details of regeneration see reference~\cite{SIAMbook}.

One small adjustment we can do to solve this problem is to first solve a slightly perturbed problem (this is the strategy employed in  \cite{hauenstReal}). Indeed, there is a nonempty Zariski open set $\Gamma$ such that for every $\gamma \in \Gamma$ the solutions to:
\begin{eqnarray}
\bS{h}(\bS{a,x,y}) - \bS{\gamma} &=& \bS{0} \label{eq:lagrangeclustersimplepert1} \\
\bS{(a,x,y)}^{T} - (a^{\star}, x^{\star}, y^{\star})^{T} &=& \bS{J}^{T}_{\bS{h}}(\bS{a,x,y})\bS{\lambda}^{T} \label{eq:lagrangeclustersimplepert2} \
\end{eqnarray}
will be nonsingular.  The benefit to first solving this system is that regeneration will now find all solutions.  After the solutions to equations \eqref{eq:lagrangeclustersimplepert1}--\eqref{eq:lagrangeclustersimplepert2} have been computed, one can use a parameter homotopy to compute all isolated solutions \eqref{eq:lagrangeclustersimple1}--\eqref{eq:lagrangeclustersimple2}, which may contain singular solutions. 

We briefly describe the parameter homotopy employed following regeneration. The solutions of equations \eqref{eq:lagrangeclustersimplepert1}--\eqref{eq:lagrangeclustersimplepert2} lead to the solutions of equations \eqref{eq:lagrangeclustersimple1}--\eqref{eq:lagrangeclustersimple2} through a collection of homotopy paths where each homotopy path as functions of $t$ are solutions to the straight-line homotopy function:
$$
\label{homotopycluster}
H(\bS{a,x,y},t) =
\begin{cases}
   \bS{h}(\bS{a,x,y}) - t \bS{\gamma} \\
\bS{(a,x,y)}^{T} - {(a^{\star}, \bS{x}^{\star}, y^{\star})}^{T} - \bS{J}^{T}_{\bS{h}}(\bS{a,x,y})\bS{\lambda}^{T}
\end{cases}
$$
for $t \in (0,1] \subset \R$. As $t \rightarrow 0$, we obtain numerical approximations to the solutions of equations \eqref{eq:lagrangeclustersimple1}--\eqref{eq:lagrangeclustersimple2}.  Additional details on parameter homotopies can be found in~\cite{SW,SIAMbook}.  A basic implementation of parameter homotopies is found in \verb!Bertini!. Input files for the regeneration and parameter homotopy runs may be found in the files {\tt Cluster\_Step1} and {\tt Cluster\_Step2}.


Timing summaries for the clustering model can be found in Table \ref{table:clusteringtimings}. In all cases, we have employed the use of intrinsically defined variables (see Appendix F.1.2 of \cite{SIAMbook}). These timings include computing the numerical irreducible decomposition of $\mathcal V_{\cM}$, the numerical irreducible decompsition of $\mathcal V_{\cM} \cap \mathcal V_{\cD}$, computing the solutions to equations 
\eqref{eq:lagrangeclustersimplepert1}--\eqref{eq:lagrangeclustersimplepert2} using regeneration, and computing the solutions to \eqref{eq:lagrangeclustersimple1}--\eqref{eq:lagrangeclustersimple2} using the parameter homotopy. We found it most appropriate to utilize \verb!Bertini! in serial for each computation except for regeneration which was done in parallel. Serial runs were done using a Apple MacBook Pro with 2.4 GHz Intel ``Core i5" processor.  Parallel runs were done using 24 (2.67 GHz Xeon-5650) compute nodes on the CentOS 5.11 operating system.
\begin{table}
\caption{Expected timings collected over 20 runs. The table includes the average time and standard deviations associated to the four computations described in this section.}
\label{table:clusteringtimings}
\begin{center}
\begin{tabular}{r|c}
    & Timing \\
\hline \hline
Compute $\mathcal V_{\cM}$ & 0.74 sec $\pm$ 0.09 sec \\
Compute $\mathcal V_{\cM} \cap \V_{\cD}$ & 0.38 sec $\pm$ 0.04 sec \\
Regeneration (parallel) & 6.09 sec $\pm$ 0.61 sec \\
Parameter Homotopy & 0.03 sec \\
\hline
\end{tabular}
\end{center}
\end{table}
In total, after reviewing the solutions, there are three solutions that correspond to real points contained in $(\mathcal V_{\cM})_{\R} \cap (\mathcal V_{\cD})_{\R}$. Among the three real solutions, two solutions are nonnegative. Solutions are listed in Table \ref{table:realsolutionscluster}. One can verify that these are indeed solutions to $(\mathcal V_{\cM})_{\R} \cap (\mathcal V_{\cD})_{\R}$ by function evaluation of $\bS{f^*}(\bS{a,x,y})$. 

We may conclude from these computations that the clustering model $\mathcal V_{\cM}$ is compatible with the observable data $\hat y$. 
\begin{table}
\caption{Nonnegative real solutions to $(\mathcal V_{\cM})_{\R} \cap (\mathcal V_{\cD})_{\R}$}
\label{table:realsolutionscluster}
\begin{center}
\begin{tabular}{l|l|ll|}
     & Solution 1 & Solution 2 \\
    \hline \hline
$\ell$ & 1.7784308 & 1.7784308 \\
$x_1$ & 0.0545838 & 0.0141547 \\
$x_2$ & 0.0952853 & 0.1357144 \\
$x_3$ & 2.1689611 & 2.1689611 \\
$\lambda$ & 1.7784308 & 1.7784308 \\
$\rho$ & 2.3188302 & 2.3188302 \\
$\zeta$ & 2.1689611 & 2.1689611 \\
$k_{o}$ & 5.3966315 & 0.1924532 \\
$k_{c}$ & 3.0914404 & 3.2734881 \\
$k_{u}$ & 3.9540082 & 0.2856796 \\
$k_s^{(2)}$ & 1.9881072 & 1.2768451 \\
$k_s^{(3)}$ & 7.6931353 & 6.9985113 \\
$k_l^{(2)}$ &1.9131209 & 1.8363315 \\
$k_l^{(3)}$ &3.6997123 & 3.6848663 \\
\hline
\end{tabular}
\end{center}
\end{table}
It may be the case that there is a very large number of data points, $\hat y$, for which we would like to determine model compatibility.  Fortunately, by employing a parameter homotopy scheme we may solve this problem rapidly where given each data point, the computations will be on the same order as the parameter homotopy solve in Table \ref{table:clusteringtimings}.

In summary the steps for model compatibility for the clustering model are as follows:
\begin{enumerate}
\item Determine the dimension of $\mathcal V_{\cM} \cap \mathcal V_{\cD}$ using the NID. 
\item Using the information gathered in Step 1, select a random point whose coordinates are sampled uniformly among a nonnegative closed interval and set up equations \eqref{eq:lagrangeclustersimplepert1}--\eqref{eq:lagrangeclustersimplepert2}. 
\item Solve the perturbed equations \eqref{eq:lagrangeclustersimplepert1}--\eqref{eq:lagrangeclustersimplepert2} from Step 2 using a regeneration scheme, for example by setting USEREGENERATION to 1 in {\tt Bertini}. 
\item Solve equations \eqref{eq:lagrangeclustersimple1}--\eqref{eq:lagrangeclustersimple2} using a parameter homotopy. Starting solutions are among the solutions gathered in Step 3.  
\item Filter the solutions gathered in Step 4 to determine if there are nonegative real solutions. 
\item We conclude that the model variety $\mathcal V_{\cM}$ is compatible with the data since there is at least one solution found in Step 5.
\end{enumerate}

\subsection{Synthetic biology and experimental design} 

We demonstrate an example from synthetic biology with excess intersection ($\dim(\mathcal{V_M}\cap\mathcal{V_D})>0$). We compare three bistable bio-circuits models analyzed in \cite{Siegal:arxiv:2014}:  monomer-dimer toggle circuit ($\mathcal M_1$), dimer-dimer  toggle circuit ($\mathcal M_2$), and single operator gene circuit ($\mathcal M_3$), which were initially presented in \cite{Gardner:2000:nature,Hasty:chaos:2001,Siegal:PLoSCB:2011}. The model variables include concentrations of genes ($X_i$) and proteins ($P_i$) where $i = 1,2$ as well as species complexes of the form $X_jP_iP_i, P_iP_i$. 

We follow the same notation for variables and parameters as presented by \cite{Siegal:arxiv:2014}. The reactions governing each of the models are given in Table~\ref{tab:reactions_synbio}.

        \begin{table}
            \caption{Each reaction described highlights whether the reaction is a forward or reversible reaction by the arrows. Here $i = 1,2$. }
        \label{tab:reactions_synbio}
        \begin{center}
        \begin{tabular}{|ccc|}
            \hline
            monomer-dimer ($\mathcal M_1$) & dimer-dimer ($\mathcal M_2$) & single operator ($\mathcal M_3$) \\
            \hline\hline
            $X_i \cee{->[k_{basi}]} X_i+P_i$ & $X_i \cee{->[k_{basi}]} X_i+P_i$ & $X_i \cee{->[k_{basi}]} X_i+P_i$\\
            $P_i \cee{->[k_{degi}]} \emptyset$ & $P_i \cee{->[k_{degi}]} \emptyset$ & $P_i \cee{->[k_{degi}]} \emptyset$\\
            $2P_2 \cee{<=>[k_{kF}][k_{kR}]} P_2P_2$ & $2P_2 \cee{<=>[k_{kF}][k_{kR}]} P_2P_2$& $2P_2 \cee{<=>[k_{kF}][k_{kR}]} P_2P_2$\\
	$X_1+P_2P_2 \cee{<=>[k_{nF}][k_{nR}]} X_1P_2P_2$ & $X_1+P_2P_2 \cee{<=>[k_{nF}][k_{nR}]} X_1P_2P_2$  & $X_2+P_2P_2 \cee{<=>[k_{qF}][k_{qR}]} X_2P_2P_2$ \\
	$X_2+P_1 \cee{<=>[k_{cF}][k_{cR}]} X_2P_1$ & $X_2+P_1P_1 \cee{<=>[k_{oF}][k_{oR}]} X_2P_1P_1$  & $X_2P_2P_2 \cee{->[k_{w}]} X_2P_2P_2+P_2$ \\
	 & $2P_1 \cee{<=>[k_{\iota F}][k_{\iota R}]} P_1P_1$  &  \\
            \hline
        \end{tabular}
        \end{center}
    \end{table}

These variables interact following mass-action kinetics and form systems of polynomial differential equations where $\mathcal M_1, \mathcal M_2,$ and $\mathcal M_3$ have 7, 8 and 6 model variables, respectively, and 10, 12, and 9 kinetic parameters, respectively. The models can be reduced by assuming that the total amount of gene 1 ($X_{1_{tot}}$) and gene 2 ($X_{2_{tot}}$) is conserved and these polynomial systems for each model are as follows. For simplicity, we use $P_{11}$ and $P_{22}$ for $P_1P_1$ and $P_2P_2$, respectively. 
\par The monomer-dimer toggle circuit ($\mathcal M_1$) system is:
\begin{align*}
-k_{deg1}P_1-k_{cF}X_2P_1+k_{bas1}X_1 + k_{cR}(X_{2_{tot}}-X_2) &=0\\
-2k_{kF}P_2^2-k_{deg2}P_2+2k_{kR}P_{22}+k_{bas2}X_2 &= 0\\
k_{kF}P_2^2-k_{kR}P_{22}-k_{nF}P_{22}X_1+k_{nR}(X_{1_{tot}} - X_1) & = 0\\
k_{nR}(X_{1_{tot}} - X_1)-k_{nF}P_{22}X_1 & = 0 \\
k_{cR}(X_{2_{tot}}-X_2)-k_{cF}P_1X_2 & = 0. 
\end{align*}
\par 
The model dimer-dimer toggle circuit ($\mathcal M_2$) system is:
\begin{align*}
-2k_{iF}P_1^2-k_{deg_1}P_1+2k_{iR}P_{11}+k_{bas1}X_1 &=0\\
k_{iF}P_1^2-k_{iR}P_{11}-k_{oF}P_{11}X_2+k_{oR}(X_{2_{tot}}-X_2) &= 0 \\
-2k_{kF}P_2^2-k_{deg2}P_2+2k_{kR}P_{22}+k_{bas2}X_2 &=0\\
k_{kF}P_2^2-k_{kR}P_{22}+k_{nF}P_{22}X_1+k_{nR}(X_{1_{tot}}-X_1) &= 0\\
k_{nR}(X_{1_{tot}} - X_1) - k_{nF}P_{22}X_1 &=0 \\
k_{oR}(X_{2_{tot}} - X_2)-k_{oF}P_{11}X_2 & = 0.
\end{align*}
The model single-operator positive feedback circuit ($\mathcal M_3$) system is:
\begin{align*}
k_{bas2}X_2-k_{deg2}P_2-2k_{kF}P_2^2+2k_{kR}P_{22}+k_w(X_{2_{tot}}-X_2) &=0\\
k_{kF}P_2^2-k_{kR}P_{22}-k_{qF}P_{22}X_2+k_{qR}(X_{2_{tot}}-X_2) &=0\\
k_{qR}(X_{2_{tot}}-X_2)-k_{qF}P_{22}X_2 & = 0
\end{align*}

In this example, we suppose that the total amounts $X_{1_{tot}}$ and $X_{2_{tot}}$ and specific protein synthesis and degradation parameters $k_{bas_1}$, $k_{bas_2}$, $k_{deg_1}$, and $k_{deg_2}$ are known and we assume that our data are measurements of $P_1$, $P_2$, and their complexes $P_{11}$, and $P_{22}$, i.e. $y=(P_1, P_2, P_{11}, P_{22})$. The aim is to select the best model $\mathcal M_1$, $\mathcal M_2$, and $\mathcal M_3$ given the data.  We simulate steady-state data $(P_1, P_2, P_{11}, P_{22})=(0.4224, 2.4153, 0.9022, 0.4758)$ from the dimer-dimer toggle model ($\mathcal M_2$) using the following parameter and variable values:

\begin{center}
\begin{tabular}{ll|ll}
    parameter & value & parameter & value \\
    \hline
    $X_{1_{tot}}$ & 1.2099 & $k_{nF}$ & 1.3566    \\
    $X_{2_{tot}}$ & 2.0660 &  $k_{nR}$ & 0.6521 \\
    $k_{bas_1}$ & 0.8718 & $k_{oF}$ & 1.5169 \\
    $k_{bas_2}$ & 1.6930 & $k_{oR}$ & 1.0661 \\
    $k_{deg_1}$ & 1.2550 & $k_{iF}$ & 3.3169  \\
    $k_{deg_2}$ & 0.6341 & $k_{iR}$ &   0.6559\\
    $k_{kF}$ & 0.6580 &  $k_{qF}$ &0.5057 \\
    $k_{kR}$ & 8.0681 &   $k_{qR}$ &   0.4844\\
    $k_{cF}$ & 0.4675 &  $k_{w}$ &  0.1478     \\
    $k_{cR}$ & 1.1636 & &    
\end{tabular}
\end{center}

We add Gaussian noise from $\mathcal{N}(0,0.1)$ and then find $\dim(\mathcal{V_{M_\text{i}}}\cap \mathcal{V_{D}})$
for $i=1,2,3$. We can compute the dimension of each intersection using the ${\tt dim}$ command in Macaulay2 or by computing a numerical irreducible decomposition in {\tt Bertini}; we find:
\begin{align*}
\dim(\mathcal{V_{M_\text{1}}}\cap(\mathcal{V_{D}})&=3, \\
\dim( \mathcal{V_{M_\text{2}}} \cap \mathcal{V_{D}})&=4, \\
\dim( \mathcal{V_{M_\text{3}}}\cap(\mathcal{V_{D}})&=3.
\end{align*}
Some further computations are required to find $\dim ((\mathcal{V_{M_\text{i}}})_{\R}\cap (\mathcal{V_{D}})_{\R}$. Specifically, we need to find real points in each intersection and determine whether or not those points are smooth.  Computing the dimension of the real part of the intersections is more work than necessary for Algorithm 2, however, it provides an illustrative example on how to work with real varieties and the algorithm in \cite{hauenstReal}.

Let $f^{(i)}=0$ be the polynomial system defining $\mathcal{V_{M_\text{i}}}\cap \mathcal{V_{D}}$ for $i=1,2,3$ and let $w^{(1)} \in \R^{17}$, $w^{(2)} \in \R^{20}$, $w^{(3)} \in \R^{15}$ be random points.  Let $x^{(i)}$ be the vector of indeterminates (unknown parameters and variables) for $i$th model, and let $c_i$ be the codimension of $\mathcal{V_{M_\text{i}}}\cap \mathcal{V_{D}}$. We can find a real point on every component of each interesection, by solving the system:

\begin{align}
&f^{(i)}=0, \label{eq:realsystem1}\\
&\lambda_1 \nabla f^{(i)}_1 + \ldots \lambda_{c_i} \nabla f^{(i)}_{c_i} + (x^{(i)}-w^{(i)}) =0 \label{eq:realsystem2}.
\end{align}

This is a simplified version of the system in Theorem 5 from \cite{hauenstReal}.  As a remark, notice the similarity of the \eqref{eq:realsystem1}-\eqref{eq:realsystem2} to the system in Theorem 1.  Algorithms for finding real points on a variety have been built on algorithms for minimizing the distance between a point and a variety since \cite{Seidenberg:1954}.

Once we have a real point on every component of $\mathcal{V_{M_\text{i}}}\cap \mathcal{V_{D}}$, we can quickly determine $\dim (\mathcal{V_{M_\text{i}}})_{\R}\cap (\mathcal{V_{D}})_{\R}$ if those real points are smooth.  Indeed, if $\mathcal V$ is an irreducible variety, then $\dim \mathcal V=\dim \mathcal V_{\R}$ if $\mathcal{V}$ contains a real smooth point (see \cite[\S 14.1]{SIAMbook}). Checking whether a real point is smooth can be done by evaluating the Jacobian $\mathcal{V_{M_\text{i}}}\cap \mathcal{V_{D}}$ at the point; if the Jacobian has full rank, then the point is smooth.  In our case, for the three models, every point we find is smooth and thus we are able to reach the conclusion:

\begin{align*}
\dim((\mathcal{V_{M_\text{1}}})_{\mathbb R}\cap(\mathcal{V_{D}})_{\mathbb R})&=3, \\
\dim( (\mathcal{V_{M_\text{2}}})_{\mathbb R}\cap(\mathcal{V_{D}})_{\mathbb R})&=4, \\
\dim( (\mathcal{V_{M_\text{3}}})_{\mathbb R}\cap(\mathcal{V_{D}})_{\mathbb R})&=3.
\end{align*}

\medskip

The dimension analysis of the varieties $\V_{\mathcal M_i} \cap \V_{\mathcal D}$ informs us about the minimum number of additional variable and parameter values that must be measured to ensure $\mathcal{V_M}\cap\mathcal{V_D}=\emptyset$.  For $\mathcal M_1$ we need to know at least 4 more variable and/or parameter values, for $\mathcal M_2$ we need to know at least 5 more, and for $\mathcal M_3$ we need to know at least 4 more.  Thus for the remainder of the example, we assume that we the parameters $k_{cF}$, $k_{cR}$, $k_{nF}$, and $k_{kF}$ are known in $\mathcal M_1$, the parameters $k_{kF}$, $k_{nF}$, $k_{iF}$, $k_{oF}$, and $k_{oR}$ are known in $\mathcal M_2$, and the parameters $k_{kF}$, $k_{qF}$, $k_{qR}$ and $k_{w}$ are known in $\mathcal M_3$.  The model $\mathcal M_3$ is an example where the number of additional parameters and/or variables that need to be known/measured exceeds the amount predicted by the dimension analysis.

Now that all the intersections are empty, we run Algorithm 2, using the regeneration methods in Bertini to solve the systems resulting from Theorem 1. We find that the sum of squares (Eq.~\eqref{eq:modelCompMin}) for each model are as follows:  $d_1^2= 2.116$, $d_2^2=0.000124$, and $d_3^2=0.6333$.  Therefore, we select the $\mathcal M_2$ model, which matches the model from which the data was generated.

Solving the zero-dimensional system for the monomer-dimer toggle circuit, $\mathcal M_1$, took 1 minute and 3 seconds on an Apple MacBook Pro with a 2.6 GhHz Intel Core i5 processor. Solving the system for the dimer-dimer toggle circuit, $\mathcal M_2$, took 1 minute and 43 seconds, and solving the system for the single-operator positive feedback circuit, $\mathcal M_3$ took 0.092 seconds.

\subsection{Epidemiology HIV} \label{subsec:HIV}


To demonstrate parameter estimation we use  a model that includes long-term HIV dynamics from initial virus, latency, and virus increase \cite{Mehta}, based on \cite{HCV07}.  Model variables $x$ are uninfected CD4+T cells ($T$), infected CD4$^+$ T cells ($T_i$), uninfected macrophages ($M$), infected macrophages ($M_i$), and HIV virus population ($V$).  The reactions are considered for this model are shown in Table \ref{tab:reactions_hiv}.

        \begin{table}
            \caption{Reactions for HIV model. The published parameter value is used from \cite{Mehta}, see references therein. }
        \label{tab:reactions_hiv}
        \begin{center}
        \begin{tabular}{|lcc|}
            \hline
            Description & Reaction & Parameter value \\
            \hline\hline
            Generation of new CD4+T cells & $\emptyset \cee{->[s_1]} T$ &  10\\
            Generation of new macrophages &$\emptyset \cee{->[s_2]} M$ &  1.5 $\times~10^{-1}$\\
            Proliferation of T cells by presence of pathogen & $T + V \cee{->[k_{1}]} (T+V)+T$ & 2 $\times$ $10^{-3}$\\
            Infection of T cells by HIV & $T + V \cee{->[k_{2}]} T_i$ & 3 $\times$ $10^{-3}$\\
            Proliferation of M by presence of pathogen & $M + V \cee{->[k_{3}]} (M+V)+M$ & 7.45 $\times$ $10^{-4}$\\
            Infection of M by HIV & $M + V \cee{->[k_{4}]} M_i$ & 5.22 $\times$ $10^{-4}$\\
            Proliferation of HIV within CD4+T cell & $T_i \cee{->[k_{5}]} V+T_i$ & 5.37 $\times$ $10^{-1}$\\
           Proliferation of HIV within macrophage & $M_i \cee{->[k_{6}]} V+M_i$ & 2.85 $\times$ $10^{-1}$\\
           Natural death of CD4+T cells & $T \cee{->[\delta_{1}]} \emptyset$ & 0.01\\
           Natural death of infected T cells  & $T_i \cee{->[\delta_{2}]} \emptyset$ & 0.44\\
           Natural death of macrophages  & $M \cee{->[\delta_{3}]} \emptyset$ & 6.6 $\times~10^{-3}$\\
           Natural death of infected macrophages  & $M_i \cee{->[\delta_{4}]} \emptyset$ & 6.6 $\times~10^{-3}$\\
           Natural death of HIV  & $V \cee{->[\delta_{5}]} \emptyset$ & 3\\
            \hline
        \end{tabular}
        \end{center}
    \end{table}

From these reactions, the dynamics are described by the following equations:
\begin{align*}
\dot T & =s_1 +k_1TV - k_2TV - \delta_1 T \\
\dot T_i & = k_2TV - \delta_2 T_i \\
\dot M  & = s_2 +k_3MV - k_4MV - \delta_3 M \\
\dot M_i & = k_4MV - \delta_4M_i \\
\dot V & = k_5T_i + k_6M_i - \delta_5V
\end{align*}

Using {\tt Macaulay2}, we find that the model variety $\V_{\mathcal M}$ has two irreducible components, the main component $\mathcal V_1$ defined by
the ideal
 \begin{align*}
 & I_1=\langle 5742M-2453M_i-130500, 259908T_i-46607M_i+4840000\delta_5-20200500, \\
 &17721T+46607M_i-4840000\delta_5+2479500, 484000V\delta_5-184547M_i+4840000\delta_5-20200500, \\ 
 &2453M_iV-72600M_i+130500V \rangle 
 \end{align*}
 and an extinction component $V_2$ defined by the ideal
 $$I_2 = \langle V, M_i,11M-250,T_i,T-1000 \rangle $$

We estimated the natural death of the virus, parameter $\delta_5$, using Algorithm 3 with main component $\mathcal V_1$ in place of $\mathcal V_{\mathcal M}$ (see Section \ref{sec:extinction}).  For $\mathcal V_{\mathcal D}$, we used the long-term non-progressors steady-state value (Table 3 of \cite{Mehta}) and added noise to each variable $\sim \mathcal{N}(0,1)$.  In particular, the data variety $\mathcal V_{\mathcal D}$ is defined by the equations:
\begin{align*}
T-\frac{6383}{20} &= 0\\
T_i - \frac{937}{20} & = 0\\
M - \frac{8109}{100} & = 0 \\
M_i - \frac{13667}{100} & = 0 \\
V - \frac{2121}{100} & = 0
\end{align*}
For $s_1, s_2, k_1, \ldots, k_6, \delta_1, \ldots, \delta_4$, we treated these parameters as known using the values from Table 1 of \cite{Mehta}.  
 
The varieties $\mathcal V_1$ and $\mathcal V_{\mathcal D}$ do not intersect, which we can confirm with {\tt Macaulay2}.  Using {\tt Bertini}, we solve the following system from Theorem 1:
\begin{align*}
& 5742M-2453M_i-130500 =0 \\
&259908T_i-46607M_i+4840000\delta_5-20200500 = 0\\
& 17721T+46607M_i-4840000\delta_5+2479500, 484000V\delta_5-184547M_i+4840000\delta_5-20200500 =0\\
& 2453M_iV-72600M_i+130500V =0\\ 
& T+17721\lambda_3-6383/20 =0\\
& T_i+259908\lambda_2-937/20  =0 \\ 
& M+5742\lambda_1-8109/100  = 0 \\
& 2453\lambda_5+M_i-2453\lambda_1-46607\lambda_2+46607\lambda_3-184547\lambda_4-72600\lambda_5-13667/100 = 0 \\
&484000\delta_5\lambda_4+2453M_i\lambda_5+V+130500\lambda_5-2121/100 =0 \\
 &     484000V\lambda_4+4840000\lambda_2-4840000\lambda_3+4840000\lambda_4 =0
\end{align*}
There are 16 complex solutions to this equation, 3 of which are real.  The real point resulting in the smallest sum of squared errors $d^2=0.2884$ is: 
\begin{align*}
&(T, T_i, M, M_i, V, \delta_5, \lambda_1, \lambda_2, \lambda_3, \lambda_4, \lambda_5) = \\
& (319.408, 46.404, 81.1544, 136.767, 21.3079, 2.99876, \\ &-0.0000112074, 0.00000171594, -0.0000145481, -0.00000519486, 0.0000159701)
\end{align*}
Thus, Algorithm 3 returns $\bar \delta_5=2.99876$, which we can compare to the true value $\delta_5 = 3$.

\subsection{Multisite phosphorylation}\label{sec:MAPK}

Here we describe the details for the multisite phosphorylation model
selection and parameter estimation computations. 
First we describe the relevant biology, next we present the mathematical models of the distributive and processive mechanisms, then we apply our model selection method using data from \cite{Aoki:2011ji}. We also
estimate the relationship between the EGF concentration and activation
of the pathway described by the parameter $k_1$ (see Table~\ref{tab:param_mapk2}).

\subsubsection{Biology of MAP Kinase system}\label{section:MAPKAoki}
Many cellular decisions are governed by molecular post-translational modifications.
One type of modification, phosphorylation, is the addition of a
phosphate group to a site of a substrate by an enzyme called a kinase.
Some proteins (substrates) require multiple phosphate groups to be
added by the kinase before the protein in activated/de-activated by
these modifications. One well-studied signaling system is the MAP
Kinase pathway, with kinase MEK and its substrate ERK; however the
mechanism by which the phosphate group is added has been debated.
Either MEK could phosphorylate ERK, disassociate and then
phosphorylate again, called {\it distributive}; or MEK could bind and
phosphorylate in sequence, called {\it processive}. Aoki et al \cite{Aoki:2011ji}
showed experimentally (with mathematical models) that this mechanism
is different {\it in vitro} than {\it in vivo}. This experiment
included 12 different levels of EGF stimulus ranging from 0.0244140625
ng/mL to 50 ng/ML. EGF actives cRAF which then phosphorylates MEK and
finally doubly phosphorylates ERK. The data are measurements of three
replicates of nonphosphorylated ERK (np-ERK), tyrosine
monophosphorylated ERK (pY-ERK), and doubly phosphorylated ERK
(pTpY-ERK), at each stimulus level.  These data are given as
percentage of total ERK (ERK), so we use the concentration
measurement for each of these ERK states.

\subsubsection{Mathematical models}
The model variables and parameters are given in Table~\ref{tab:description_mapk}. 
The model parameters for the distributive model are $${\bS{a}}=(k_1, \ldots, k_{27}, c_1, c_2), $$  
the variables are $$\bS{x} = (x_1, \ldots, x_{12}, \text{cRAF}_{tot}, \text{MEK}_{tot}, \text{ERK}_{tot}),$$
and the outputs are
$$y=(\text{np-ERK},\text{pY-ERK},\text{pYpT-ERK}).$$ 
The variables for the processive model are the same as for the distributive model except for two additional variables $x_{13}, x_{14}$. The reaction velocities are given in 
Table~\ref{tab:reaction_mapk} and the corresponding equations are given in Table~\ref{tab:eqs_mapk}.  Note in Table~\ref{tab:eqs_mapk} that there are various conserved species concentrations in addition to the ODEs.
 \begin{table}
            \caption{Description of variables and parameters for distributive and processive MAP Kinase models }
        \label{tab:description_mapk}
        \begin{center}
\begin{tabular}{|ll|llll|}
\hline
    variable & species & parameter & name & parameter & name\\
    \hline \hline
    $x_1$ & MEK & $k_1$  & kphos\_MEK\_pMEK &$k_{15}$ & kdphos\_pY\_np\_cyt \\
    $x_2$ & cRAF & $k_2$  & kdphos\_pMEK\_MEK &$k_{16}$ & kdphos\_pT\_np\_cyt \\
    $x_3$ & pMEK & $k_3$ & kf\_MEK\_ERK\_binding&$k_{17}$& kdphos\_pTpY\_pY\_nuc \\
    $x_4$ & np-ERK\_cyt & $k_4$  & kb\_MEK\_ERK\_dissociation& $k_{18}$ & kdphos\_pTpY\_pT\_nuc \\
    $x_5$ & MEK\_np-ERK & $k_5$ & kimport\_np &$k_{19}$ & kdphos\_pY\_np\_nuc \\
    $x_6$ & np-ERK\_nuc & $k_6$  & kexport\_np &$k_{20}$ & kdphos\_pT\_np\_nuc \\
    $x_7$ & pY-ERK\_cyt & $k_7$  & kimport\_pY &$k_{21}$ & kphos\_np\_pY \\
    $x_8$ & pY-ERK\_nuc & $k_8$  & kexport\_pY &$k_{22}$ & kphos\_pY\_pTpY \\
    $x_9$ & pT-ERK\_cyt & $k_9$  & kimport\_pT&$k_{23}$ & kphos\_pT\_pTpY \\
    $x_{10}$ & pT-ERK\_nuc & $k_{10}$ & kexport\_pT &$k_{24}$ & kf\_MEK\_ERK\_binding \\
    $x_{11}$ & pTpY-ERK\_cyt & $k_{11}$ & kimport\_pTpY &$k_{25}$ & kb\_MEK\_ERK\_dissociation \\
    $x_{12}$ & pTpY-ERK\_nuc & $k_{12}$ & kexport\_pTpY &$k_{26}$ & kphos\_np\_pY \\
    $x_{13}$ & pMEK\_np-ERK & $k_{13}$ & kdphos\_pTpY\_pY\_cyt& $k_{27}$ & kphos\_pY\_pTpY\_MEKERK \\
    $x_{14}$ & pMEK\_pY-ERK & $k_{14}$ & kdphos\_pTpY\_pT\_cyt&$c_{2}$,$c_{1}$  & cyt\_vol, nuc\_vol  \\
    \hline
    \end{tabular}
        \end{center}
    \end{table}
\begin{table}
            \caption{Reaction velocities for the MAP Kinase distributive and processive model.  The processive model uses the additional reaction velocities $v_{18},v_{19},v_{20}$.}
        \label{tab:reaction_mapk}
        \begin{center}
\begin{tabular}{|lll|}
\hline
    $v_1 = k_1 x_1 x_2 - k_2 x_3$ &  $ v_2 = k_3 x_1 x_4 - k_4 x_5$ &   $v_3 = k_5 x_4 - c_2 k_6 x_6$ \\
 $  v_4 = k_7 x_7 - c_2 k_8 x_8$ &  $v_5 =k_9 x_9 - c_2 k_{10} x_{10}$ &   $v_6 = k_{11} x_{11} - c_2 k_{12} x_{12} $\\
   $v_7 = k_{13} x_{11}$  &    $v_8 = k_{14} x_{11} $&  $v_9 = k_{15} x_7  $\\
  $ v_{10} = k_{16} x_{9}$ &    $ v_{11} =  c_2 k_{17} x_{12}$ &    $ v_{12} = c_2 k_{18} x_{12} $\\
   $ v_{13} = c_2 k_{19} x_8 $ & $ v_{14} = c_2 k_{20} x_{10}$ & $    v_{15} = k_{21} x_3 x_4 $\\
    $v_{16} = k_{22} x_3 x_7$ &    $v_{17} = k_{23} x_3 x_9$ & \\ \hline
 $v_{18} = k_{24}x_{3}x_{4} - k_{25}x_{13}$ & $v_{19} = k_{26}x_{13}$ & $v_{20} = k_{27}x_{14}$\\
\hline
    \end{tabular}
        \end{center}
    \end{table}

 \begin{table}
            \caption{Equations for distributive and processive MAP Kinase models }
        \label{tab:eqs_mapk}
        \begin{center}
\begin{tabular}{|rl|l|l|}
\hline
    Variable & & Distributive & Processive  \\
    \hline \hline
    $\dot{x}_1$&$=$ & $-v_1 - v_2$ &$-v_1 - v_2$ \\
    $\dot{x}_2$&$=$ & 0 & 0 \\
    $\dot{x}_3$&$=$ & $v_1$ & $v_1 - v_{18} + v_{20}$\\
    $\dot{x}_4$&$=$ & $-v_2 - v_3 + v_9 + v_{10} - v_{15}$ &$-v_2 - v_3 + v_9 + v_{10} - v_{18}$\\
    $\dot{x}_5$&$=$ & $v_2$ & $v_2$\\
    $\dot{x}_6$&$=$ & $v_3 + v_{13} + v_{14}$ & $v_3 + v_{13} + v_{14}$\\
    $\dot{x}_7$&$=$ & $-v_4 + v_7 - v_9 + v_{15} - v_{16}$ & $-v_4 + v_7 - v_9 - v_{16}$\\
    $\dot{x}_8$&$=$ & $v_4 + v_{11} - v_{13}$  & $v_4 + v_{11} - v_{13}$ \\
    $\dot{x}_9$&$=$ & $-v_5 + v_8 - v_{10} - v_{17}$&$-v_5 + v_8 - v_{10} - v_{17}$\\
    $\dot{x}_{10}$&$=$ & $v_5 + v_{12} - v_{14}$& $v_5 + v_{12} - v_{14}$ \\
    $\dot{x}_{11}$&$=$ & $-v_6 - v_7 - v_8 + v_{16} + v_{17}$ &$-v_6 - v_7 - v_8 + v_{16} + v_{17}+v_{20}$ \\
    $\dot{x}_{12}$&$=$ & $v_6 - v_{11} - v_{12}$ & $v_6 - v_{11} - v_{12}$ \\
    $\dot{x}_{13}$&$=$ &  & $v_{18} - v_{19}$\\
    $\dot{x}_{14}$&$=$ &  & $v_{19} - v_{20}$\\
    $0$&$=$ & MEK$_{tot} - (x_1 + x_3 + x_5)$ & MEK$_{tot} - (x_1 + x_3 + x_5 + x_{13} + x_{14})$ \\
    $0$&$=$ & cRAF$_{tot} - x_2$  &  cRAF$_{tot} - x_2$\\
    $0$&$=$ & ERK$_{tot} - \sum_{i=4}^{12} x_{i}$ & ERK$_{tot} - \sum_{i=4}^{14} x_{i}$ \\
    \hline
    \end{tabular}
        \end{center}
    \end{table}
%

%
We use the \emph{in vitro}  parameters estimates from Table S2 in reference \cite{Aoki:2011ji} for $k_2, \ldots, k_{27}, c_1, c_2$ and the conserved quantities MEK$_{tot}$, cRAF$_{tot}$, ERK$_{tot}$, as given in Table~\ref{tab:param_mapk}. The remaining parameter, $k_1$, describes the rate of MEK phosphorylation and depends on the level of EGF stimulation, which varies throughout the data. Thus, we left it as a free variable and estimated it as a byproduct of distance minimization (\ref{tab:param_mapk2}).
 \begin{table}
            \caption{Parameter values for MAP Kinase models}
        \label{tab:param_mapk}
\begin{center}
\begin{tabular}{|ll|ll|ll|}
\hline
    parameter & value & parameter & value & parameter & value \\
    \hline \hline
    $k_2$ & 0.0096 & $k_{13}$ & 0.004  & $k_{24}$ & 0.18  \\
    $k_3$ & 0.18 &  $k_{14}$ & 0.0055 & $k_{25}$ & 0.27\\
    $k_4$ & 0.27 & $k_{15}$ & 0.0067 & $k_{26}$ & 0.073\\
    $k_5$ & 0.0017 & $k_{16}$ & 0.0068 & $k_{27}$ & 0.05 \\
    $k_6$ & 0.013 & $k_{17}$ & 0.0032 & $c_1$ & 1.0 \\
    $k_7$ & 0.0025 & $k_{18}$ & 0.0038 & $c_2$ & 0.2 \\
    $k_8$ & 0.017 & $k_{19}$ & 0.0077 & cRAF$_{tot}$ & 0.013 \\
    $k_9$ & 0.0022 & $k_{20}$ & 0.0058 &  MEK$_{tot}$ & 1.2 \\
    $k_{10}$ & 0.049 &  $k_{21}$ & 0.039  & ERK$_{tot}$ & 0.74 \\
    $k_{11}$ & 0.0082 & $k_{22}$ & 0.021 && \\
   $k_{12}$ & 0.0076  &  $k_{23}$ & 0.02 && \\
   \hline
\end{tabular}
 \end{center}
    \end{table}
The output variables are  np-ERK, pY-ERK, and pYpT-ERK, which are sums of species concentrations.
For the distributive model, the output equations are:
\begin{eqnarray}
\text{np-ERK} - (x_4 + x_5 + x_6) &=& 0 \label{eq:sum1}\\
\text{pY-ERK} - (x_7 + x_8) & = & 0\\
\text{pYpT-ERK} - (x_{11} + x_{12}) & = & 0 \label{eq:sum3}
\end{eqnarray}
whereas for the processive model, we include two additional species, so the output equations become:
\begin{eqnarray}
\text{np-ERK} - (x_4 + x_5 + x_6+x_{13}) &=& 0 \label{eq:sum1b}\\
\text{pY-ERK} - (x_7 + x_8+x_{14}) & = & 0\\
\text{pYpT-ERK} - (x_{11} + x_{12}) & = & 0. \label{eq:sum3b}
\end{eqnarray}



\subsubsection{Model selection and parameter estimation computations} \label{subsect:MAPKopt}
The model variety  $\mathcal V_{\cM_d}$ of the distributive model is defined by \eqref{eq:sum1} - \eqref{eq:sum3} and the equations obtained by setting the ``Distributive" column of Table \ref{tab:eqs_mapk} equal to zero.  We will refer to the system defining $\mathcal V_{\cM_d}$ as $F=0$. The model variety  $\mathcal V_{\cM_p}$ of the processive model is defined by \eqref{eq:sum1b} - \eqref{eq:sum3b} and the equations obtained by setting the ``Processive" column of Table \ref{tab:eqs_mapk} equal to zero.  The ambient dimension of $\mathcal V_{\cM_d}$ is $16$ since the coordinates that define $\mathcal V_{\cM_d}$ include $x_1, \dots, x_{12}$, np-ERK, pY-ERK, pYpT-ERK, and the model parameter $k_{1}$; all other parameters and variables we treat as known constants.  Similarly the ambient dimension for the processive model is $18$ as we include the additional variables $x_{13}$ and $x_{14}$.


Given data $\hat y= (\widehat{\text{np-ERK}}, \widehat{\text{pY-ERK}}, \widehat{\text{pYpT-ERK}})$ we define the data variety for the distributive model as:

$$\mathcal V_{\mathcal{D}_d} = \left\{\bS{(a, x, y)} \in \C^{16} : y = \hat y \right\}$$


The data variety $\mathcal V_{\mathcal{D}_d}$ has dimension $13$. The specific data that we used may be found in the supplementary data file ({\tt aoki\_data.txt}).  The data variety $\mathcal V_{\cD_p}$ for the processive model is defined similarly.

The computations that follow will be for the distributive model.  The computations for the processive model will be nearly identical so we do not describe them in the same level of detail.  When the results are discussed we will be sure to record information for both models. 

We first compute a numerical irreducible decomposition (NID) of $\mathcal V_{\mathcal{M}_d}$ using the {\tt Bertini.m2} \cite{BGLR} the {\tt Macaulay2} interface for {\tt Bertini} to solve \cite{GS,BGLR,bertini}. 
With the NID, one can verify that $\mathcal V_{\cM_d}$ is a one-dimensional complex algebraic set of degree $8$ (filename: {\tt MAPK\_D\_Model\_NID}). Similarly for the processive model, the model variety $\mathcal V_{{\cM}_p}$ is a one-dimensional complex algebraic set of degree $11$ (filename: {\tt MAPK\_P\_Model\_NID}).  There are several variables that may be intrinsically defined to save computation.  For example, $x_{1}, x_{2}, x_{7}$, and  $x_{11}$ can be written in terms of the other variables followed by $x_{4}$. One may also verify that $\mathcal V_{\mathcal{M}_d} \cap \mathcal V_{\mathcal{D}_d} = \emptyset$  using {\tt Bertini} (filename: {\tt MAPK\_D\_Model\_Data\_NID}).  Here, one can also define the variables np-ERK, pY-ERK, and pYpT-ERK intrinsically to save computation. Since $\mathcal V_{\mathcal{M}_d} \cap \mathcal V_{\mathcal{D}_d} = \emptyset$, Algorithm 2 instructs us to minimize the distance between $(\mathcal V_{\mathcal{M}_d})_{\R}$ and $(\mathcal V_{\mathcal{D}_d})_{\R}$.


Squaring up the polynomial system defining $V_{\mathcal{M}_d}$ will be a necessary step to construct the polynomial system from Proposition 1.  This procedure was described briefly in Section \ref{subsec:geometry} and in more detail in Section \ref{subsec:celldeath}.  The codimension of $V_{\mathcal{M}_d}$ is $c=16-1=15$, the dimension of the ambient space minus the dimension of $V_{\mathcal{M}_d}$ as determined by the numerical irreducible decomposition.  Let $A \in \C^{15 \times 17}$ whose entries are taken randomly from the complex unit circle.  The polynomial system from Proposition 1 becomes: 
\begin{eqnarray}
f'_{j}(\bS{a,x,y}) &=& 0, \text{ for }  1 \leq i \leq 15, \text{ where } f'_{j}(\bS{x}) = \Sigma_{k=1}^{17} A_{jk} F_{k}(\bS{a,x,y}) \text{ with } A_{jk} = [\bS{A}]_{j,k} \label{eq:local1}\\
0 &=& \Sigma_{j=1}^{15} \frac{\partial f'_{j}(\bS{a,x,y})}{\partial a_{1}}\lambda_{j}, \\
0 &=& \Sigma_{j=1}^{15} \frac{\partial f'_{j}(\bS{a,x,y})}{\partial x_{i}}\lambda_{j}, \text{ for } 1 \leq i \leq 12 \\
y_{i}-\hat y_{i} &=& \Sigma_{j=1}^{15} \frac{\partial f'_{j}(\bS{a,x,y})}{\partial y_{i}}\lambda_{j}, \text{ for } 1 \leq i \leq 3 \label{eq:local5}\
\end{eqnarray}

The variables $\bS{\lambda} = (\lambda_1,\ldots,\lambda_{15})$ are the Lagrange multipliers.  This is a system of $31$ variables and $31$ equations. 
We collect the solutions $(\bS{a, x, y, \lambda}) \in \R \! \times \R^{12} \! \times \R^3 \! \times \C^{15}$ to equations \eqref{eq:local1}--\eqref{eq:local5} and then compute $\|\bS{y} - \hat y\|_{2}^{2}$ for each solution.

In Section \ref{subset:boundary}, we explained an issue that can arise in solving constrained optimization problems such as ones arising from problem \eqref{eq:KKT1}--\eqref{eq:KKT5}. 
In this example, we want  to ensure that $x_1, \ldots, x_{12}, a_1, y_1, y_2, y_3$ are non-negative, i.e. $\mathcal S_{\mathcal{M}_d}=\mathcal V_{\mathcal{M}_d} \cap \R_{\geq 0}^{16}$.  To minimize the distance between $\mathcal S_{\mathcal{M}_d}$ and $\mathcal V_{\mathcal D_d}$ using a numerical algebraic geometry approach, we should solve the system \eqref{eq:local1}--\eqref{eq:local5}, and then solve the system again 16 more times, setting one of $x_1, \ldots, x_{12}, a_1, y_1, y_2, y_3$ to zero each time.  We know that this will be sufficient to check the boundary, since the intersection of $\mathcal V_{\mathcal{M}_d}$ with any coordinate hyperplane is zero dimensional.

The data set we used can be found in the file ({\tt aokidata.txt}). There are 36 sets of data points where each set consists of a triple $\hat y= (\widehat{\text{np-ERK}}, \widehat{\text{pY-ERK}}, \widehat{\text{pYpT-ERK}})$. Each data point defines a data variety, which we will denote $(V_{\mathcal{D}_{i}})_{\mathbb{R}}$ for the $i$th data point. 

The polynomial system \eqref{eq:local1}--\eqref{eq:local5} is a classic example of a parameterized system of polynomial equations with parameter $\hat y$ (notice that this is different than the rate parameters of the ODE).  Let $\bS{J}_{\bS{f'}}^{T}(\bS{a,x,y})$ be the Jacobian matrix of $f'(a, x, y)$. The theory for parameter homotopies states that there is a nonempty Zariski open set $U \subset \C^{3}$ such that for every $\widehat{\bS{y}}^{\star} \in U$ the nonsingular solutions of:
\begin{eqnarray}
\bS{f}'(\bS{a,x,y}) &=& \bS{0} \label{eq:parametergeneral1} \\
\begin{bmatrix}
\bS{0} \\
\bS{{y}}^{T} - (\widehat{\bS{y}}^{\star})^{T} \\
\end{bmatrix}&=& \bS{J}_{\bS{f'}}^{T}(\bS{a,x,y})\bS{\lambda}^{T} \label{eq:parametergeneral2} \
\end{eqnarray}
lead to solutions of equations \eqref{eq:local1}--\eqref{eq:local5} by homotopy paths. Each homotopy path is the set of zeros of the straight-line homotopy function of $t \in (0,1]$ as $t$ varies from 1 to 0:
$$
\label{homotopy}
H(\bS{x},t) =
\begin{cases}
   \bS{f}'(\bS{a,x,y}) \\
   \begin{bmatrix}
   {0} \\
  y^{T} - \left(t (\widehat{\bS{y}}^{\star})^{T} + (1 - t) \widehat{\bS{y}}^{T}\right) \\
   \end{bmatrix} - \bS{J}_{\bS{f'}}^{T}(\bS{a,x,y})\bS{\lambda}^{T}
\end{cases}.
$$
As $t \rightarrow 0$, we obtain numerical approximations to the solutions of equations \eqref{eq:local1}--\eqref{eq:local5}. 

The benefit of employing a parameter homotopy is that after solving equations \eqref{eq:parametergeneral1}--\eqref{eq:parametergeneral2} with a more general method \emph{a priori}, we significantly reduce the computation required in solving equations \eqref{eq:local1}--\eqref{eq:local5} 
for each $i$. In practice, the entries of $\widehat{\bS{y}}^{\star}$ are sampled uniformly along the complex unit circle. More details on how parameter homotopies fit into numerical algebraic geometry can be found in \cite{SW} and \cite{SIAMbook}. 

In addition to employing a parameter homotopy solving scheme, equations  \eqref{eq:parametergeneral1}--\eqref{eq:parametergeneral2}, or equally equations \eqref{eq:local1}--\eqref{eq:local5},  have a natural $\{\bS{(a,x,y),\lambda}\}$-homogenous structure (see \cite{SW} and \cite{SIAMbook}). This observation significantly reduces the number of homotopy paths that need to be tracked numerically. In addition, this increases stability of path tracking. Multihomogenous structures are used alongside parameter homotopies to solve equations \eqref{eq:local1}--\eqref{eq:local5}.

One additional reformulation that we can do to reduce computation is to define some of the variables intrinsically.  This is common if one or more variables can be written as a linear combination of some of the other variables.   Specifically,  we know from Table~\ref{tab:eqs_mapk} that:
$$
x_{2} = \text{cRAF}_{tot}
$$
where cRAF$_{tot}$ is defined as a constant in Table~\ref{tab:param_mapk}.  Thus, we can ``remove" $x_{2}$ from our computations. Partial derivatives are no longer necessary with respect to the variable $x_{2}$, and $x_{2}$ is no longer defined explicitly when tracking homotopy paths.

\begin{table}
\caption{Path counts on processive and distributive models. `$\{\bS{(a,x,y)},\bS{\lambda}\}$-hom' corresponds to a $\{\bS{(a,x,y},\bS{\lambda}\}$-homogeneous variable grouping and `intrinsic $x_2$' corresponds to the system where $x_2$ is intrinstically defined.}
\label{table:aokipaths}
\begin{center}
\begin{tabular}{l|c|c|c}
    & Total Degree & $\{\bS{(a,x,y)},\bS{\lambda}\}$-hom & $\{\bS{(a,x,y)},\bS{\lambda}\}$-hom $+$ intrinsic $x_2$ \\
\hline \hline
Processive Model & 124416 paths &  3744 paths & 1152 paths \\
Distributive Model & 248832 paths & 7488 paths & 2304 paths \\
\hline
\end{tabular}
\end{center}
\end{table}

Table~\ref{table:aokipaths} summarizes the sequence of reductions made in the number of paths by imposing a $\{\bS{(a,x,y)},\bS{\lambda}\}$-homogeneous structure followed by intrinsically defining the variable $x_{2}$ along with the number of paths required using the standard total degree homotopy \cite{SW},\cite{SIAMbook}.

\begin{figure}[ht]
\caption{Distance Plot}
\label{fig:distanceplot}
\begin{center}
	\includegraphics[clip=true, trim = 30mm 0 0 0, scale=.52]{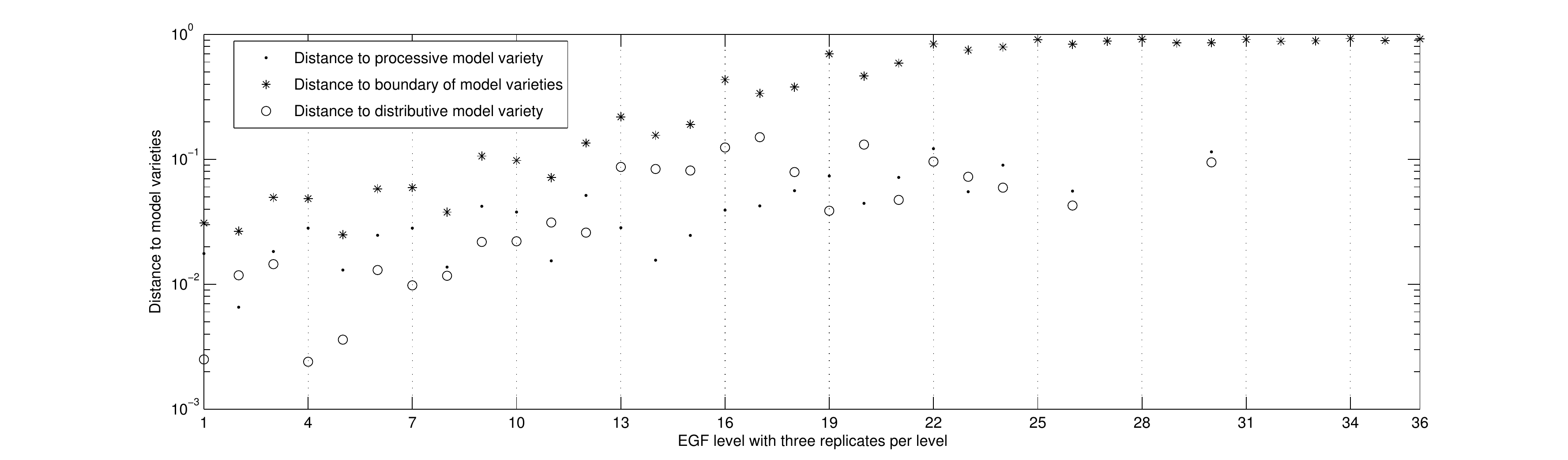}
\end{center}
\end{figure}

Table~\ref{table:smalleraokiresults} and Table~\ref{table:largeraokiresults} record the distances between the data and model varieties for all 36 data points. A missing ``interior'' distance in Table~\ref{table:smalleraokiresults} and Table~\ref{table:largeraokiresults} indicate there were no positive real critical points found for the given EGF level and replicate. However, we may still compute a distance to the boundary of the semi-algebraic sets corresponding to each model. \verb!Bertini! input files, shell scripts, and \verb!MATLAB! scripts are available within the supplementary files to analyze model selection and parameter estimation.   The distances are summarized graphically in Figure~\ref{fig:distanceplot}.

\begin{table}
\caption{Expected timings for the MAPK model collected over 20 `random' runs.}
\label{table:aokitimings}
\begin{center}
\begin{tabular}{l|c|c|c}
    & Compute Dimension & Initial Solve (parallel) & Data Solve (all 36) \\
\hline \hline
Processive Model & $9.14$ sec $ \pm 1.21$ sec & 28.05 sec $\pm$ 3.24 sec & 10.95 sec $\pm 2.09$ sec \\
Distributive Model & $13.54$ sec $\pm 1.08$ sec & 53.06 sec $\pm$ 9.20 sec & $19.09$ sec $ \pm 4.84$ sec \\
\hline
\end{tabular}
\end{center}
\end{table}

Timing summaries for both the processive and distributive model can be found in Table~\ref{table:aokitimings}.  These timings include the numerical irreducible decomposition required to compute the dimension of each component in the model variety, solving equations \eqref{eq:parametergeneral1}--\eqref{eq:parametergeneral2} required to employ a parameter homotopy scheme, and the parameter homotopy to solve equations \eqref{eq:local1}--\eqref{eq:local5} for all $i$.  Timings to compute the dimension of the model variety and the data solve were done in serial using a Apple MacBook Pro with 2.4 GHz Intel ``Core i5" processor.  The initial solve for the parameter homotopies were done in parallel using 96 (2.67 GHz Xeon-5650) compute nodes on the CentOS 5.11 operating system.

 \begin{table}[ht]
            \caption{Parameter estimate of $k_1$ for distributive and processive MAP Kinase models }
        \label{tab:param_mapk2}
\begin{center}
\begin{tabular}{l|l|l}
\hline
    EGF level & Distributive & Processive \\
    \hline \hline
    1 (0.0244140625 ng/mL) & 0.006655185015566 & 0.002630893498837  \\
    1  & 0.005169208080985 & 0.002666996926268  \\
    1 & 0.010517845922915 & 0.004869916688582  \\
    2 (0.048828125 ng/mL) & 0.010599752816972 & 0.004139244229281  \\
    2  & 0.005294859090859 & 0.002185712163548  \\
    2 & 0.012645415710605 & 0.005598240936340  \\
 3 (0.09765625 ng/mL) & 0.013040547423470 & 0.005555450037129  \\
    3  & 0.007862190037618 & 0.003676690633723  \\
    3 & 0.007862190037618 & 0.010890090940375  \\
 4 (0.1953125 ng/mL) & 0.022314241866226 & 0.007566455646161  \\
    4  & 0.014767039426564 & 0.010026925643431  \\
    4 & 0.032112677267837 & 0.014358973830327  \\
 5 (0.390625 ng/mL) & 0.057037089355901 & 0.028188983627261  \\
    5  & 0.034598433900385 & 0.018615955020805  \\
    5 & 0.046993978170041 & 0.023610675947125  \\
 6 (0.78125 ng/mL) & 0.171132616846834 & 0.081810937526556  \\
    6  & 0.108600436914432 & 0.052541291660911  \\
    6 & 0.128469450822607 & 0.062127115025643  \\
 7 (1.5625 ng/ML) & 0.552602449693322 & 0.311829951745710  \\
    7  & 0.198177806441869 & 0.094130793284512  \\
    7 & 0.307630980846653 & 0.162410456627761  \\
 8 (3.125 ng/ML) & 1.535918937663066 & 1.104298831092584  \\
    8  & 1.558792683503375 & 0.653311235583287  \\
    8 & 1.114642498639051 & 0.700052847271085  \\
 9 (6.25 ng/ML) & 0 & 0  \\
    9  & 6.741089632275663 & 2.682148283074403  \\
    9 & 0 & 0  \\
 10 (12.5 ng/ML) & 0 & 0  \\
    10  & 0 & 0  \\
    10 & 2.556601780467115 & 1.856045810178439  \\
11 (25 ng/ML) & 0 & 0  \\
    11  & 0 & 0  \\
    11 & 0 & 0  \\
 12 (50 ng/ML) & 0 & 0  \\
    12  & 0 & 0  \\
    12 & 0 & 0  \\
   \hline
\end{tabular}
 \end{center}
    \end{table}
\vspace{2pc}

\begin{table}
\caption{Distance to (smaller) distributive model variety}
\label{table:smalleraokiresults}
\begin{center}
\begin{tabular}{c|c|c|c}
    EGF level & Replicate & ``Interior'' distance & ``Boundary'' distance \\
    \hline \hline
          $1$ & $1$ & 0.0025 & 0.0309 \\
                 & $2$ & 0.0118 & 0.0266 \\
                 & $3$ & 0.0145 & 0.0496 \\
    \hline
          $2$ & $1$ & 0.0024 & 0.0485 \\
                 & $2$ & 0.0036 & 0.0249 \\
                 & $3$ & 0.0130 & 0.0581 \\
    \hline
          $4$ & $1$ & 0.0098 & 0.0594 \\
                 & $2$ & 0.0117 & 0.0377 \\
                 & $3$ & 0.0218 & 0.1062 \\
    \hline
          $8$ & $1$ & 0.0221 & 0.0981 \\
                 & $2$ & 0.0312 & 0.0714 \\
                 & $3$ & 0.0259 & 0.1349 \\
    \hline
        $16$ & $1$ & 0.0870 & 0.2189 \\
                 & $2$ & 0.0838 & 0.1559 \\
                 & $3$ & 0.0814 & 0.1904 \\
    \hline
        $32$ & $1$ & 0.1243 & 0.4343 \\
                 & $2$ & 0.1505 & 0.3374 \\
                 & $3$ & 0.0791 & 0.3784 \\
    \hline
        $64$ & $1$ & 0.0388 & 0.6990 \\
                 & $2$ & 0.1312 & 0.4648 \\
                 & $3$ & 0.0473 & 0.5889 \\
    \hline
      $128$ & $1$ & 0.0959 & 0.8398 \\
                 & $2$ & 0.0725 & 0.7501 \\
                 & $3$ & 0.0594 & 0.7931 \\
    \hline
      $256$ & $1$ & ---        & 0.9093 \\
                 & $2$ & 0.0427 & 0.8353 \\
                 & $3$ & ---            & 0.8839 \\
    \hline
      $512$ & $1$ & ---            & 0.9154 \\
                 & $2$ & ---            & 0.8556 \\
                 & $3$ & 0.0947 & 0.8597 \\
    \hline
    $1024$ & $1$ & ---            & 0.9111 \\
                 & $2$ & ---           & 0.8817 \\
                 & $3$ & ---            & 0.8883 \\
    \hline
    $2048$ & $1$ & ---            & 0.9272 \\
                 & $2$ & ---            & 0.8948 \\
                 & $3$ & ---            & 0.9197 \\
    \hline
\end{tabular}
\end{center}
\end{table}

\begin{table}
\caption{Distance to (larger) processive model variety}
\label{table:largeraokiresults}
\begin{center}
\begin{tabular}{c|c|c|c}
    EGF level & Replicate & ``Interior'' distance & ``Boundary'' distance \\
    \hline \hline
          $1$ & $1$ & 0.0176 & 0.0309 \\
                 & $2$ & 0.0066 & 0.0266 \\
                 & $3$ & 0.0183 & 0.0496 \\
    \hline
          $2$ & $1$ & 0.0281 & 0.0485 \\
                 & $2$ & 0.0130 & 0.0249 \\
                 & $3$ & 0.0247 & 0.0581 \\
    \hline
          $4$ & $1$ & 0.0282 & 0.0594 \\
                 & $2$ & 0.0137 & 0.0377 \\
                 & $3$ & 0.0421 & 0.1062 \\
    \hline
          $8$ & $1$ & 0.0379 & 0.0981 \\
                 & $2$ & 0.0154 & 0.0714 \\
                 & $3$ & 0.0514 & 0.1349 \\
    \hline
        $16$ & $1$ & 0.0284 & 0.2189 \\
                 & $2$ & 0.0156 & 0.1559 \\
                 & $3$ & 0.0246 & 0.1904 \\
    \hline
        $32$ & $1$ & 0.0392 & 0.4343 \\
                 & $2$ & 0.0424 & 0.3374 \\
                 & $3$ & 0.0561 & 0.3784 \\
    \hline
        $64$ & $1$ & 0.0735 & 0.6990 \\
                 & $2$ & 0.0444 & 0.4648 \\
                 & $3$ & 0.0717 & 0.5889 \\
    \hline
      $128$ & $1$ & 0.1218 & 0.8398 \\
                 & $2$ & 0.0550 & 0.7501 \\
                 & $3$ & 0.0899 & 0.7931 \\
    \hline
      $256$ & $1$ & ---            & 0.9093 \\
                 & $2$ & 0.0557 & 0.8353 \\
                 & $3$ & ---            & 0.8839 \\
    \hline
      $512$ & $1$ & ---            & 0.9154 \\
                 & $2$ & ---            & 0.8556 \\
                 & $3$ & 0.1149 & 0.8597  \\
    \hline
    $1024$ & $1$ & ---            & 0.9111 \\
                 & $2$ & ---            & 0.8817 \\
                 & $3$ & ---            & 0.8883 \\
    \hline
    $2048$ & $1$ & ---            & 0.9272 \\
                 & $2$ & ---            & 0.8948 \\
                 & $3$ & ---            & 0.9197 \\
    \hline
\end{tabular}
\end{center}
\end{table}

\bibliographystyle{pnas}
\bibliography{references}